\documentclass{aa}  

\usepackage{graphicx}
\usepackage{verbatim}
\usepackage{multirow}
\usepackage[utf8]{inputenc}
\usepackage{longtable}
\usepackage{txfonts}
\usepackage[colorlinks=true,allcolors=blue]{hyperref}

\newcommand{\sect}[1]{\text{Sect.~\ref{#1}}} 
\newcommand{\fig}[1]{\text{Fig.~\ref{#1}}} 

\newcommand{\marcs}{\texttt{MARCS}}
\newcommand{\phoenix}{\texttt{PHOENIX}}
\newcommand{\sme}{\texttt{SME}}

\newcommand{\vald}{\texttt{VALD}}

\newcommand{\teff}{T_{\mathrm{eff}}} 
\newcommand{\lgg}{\log{g}} 
\newcommand{\feh}{\mathrm{\left[Fe/H\right]}} 
\newcommand{\dex}{\mathrm{dex}} 
\newcommand{\kelvin}{\mathrm{K}} 

\begin{document}

   \title{Comparative high-resolution spectroscopy of M~dwarfs -- exploring non-LTE effects}


   \author{T. Olander
          \and
          U. Heiter
          \and O. Kochukhov  
          }

   \institute{Observational Astrophysics, Department of Physics and Astronomy, Uppsala University, Box 516, 75120 Uppsala, Sweden,
              \\
              \email{terese.olander@physics.uu.se,  ulrike.heiter@physics.uu.se, oleg.kochukhov@physics.uu.se}
         }

   \date{Received ; accepted }

 
  \abstract
{M~dwarfs are key targets for high-resolution spectroscopy and model atmosphere analyses due to a high incidence of these stars in the solar neighbourhood and their importance as exoplanetary hosts. Several methodological challenges make such analyses difficult, leading to significant discrepancies in the published results.}
{The aim of our work is to compare M~dwarf parameters derived by recent high-resolution near-infrared studies with each other and with fundamental stellar parameters. We also assess to what extent deviations from local thermodynamic equilibrium (LTE) for iron and potassium influence the outcome of these studies.}
{We carry out line formation calculations based on a modern model atmosphere grid appropriate for M~dwarfs along with a synthetic spectrum synthesis code that treats formation of atomic and molecular lines in cool-star atmospheres including departures from LTE. We use near-infrared spectra collected with the CRIRES instrument at the ESO VLT as reference observational data.}
{We find that the effective temperatures obtained with spectroscopic techniques in different studies mostly agree to better than 100~K and are mostly consistent with the fundamental temperatures derived from interferometric radii and bolometric fluxes. At the same time, a much worse agreement in the surface gravities and metallicities is evident. Significant discrepancies in the latter parameters appear when results of the studies based on the optical and near-infrared observations are intercompared. We demonstrate that non-LTE effects are negligible for \ion{Fe}{i} in M-dwarf atmospheres but are important for \ion{K}{i}, which has a number of strong lines in near-infrared spectra of these stars. These effects, leading to potassium abundance and metallicity corrections on the order of 0.2~dex, may be responsible for some of the discrepancies in the published analyses. Differences in the temperature-pressure structures of the atmospheric models may be another factor contributing to the deviations between the spectroscopic studies, in particular at low metallicities and high effective temperatures.}
{High-resolution spectroscopic studies of M~dwarfs are yet to reach the level of consistency and reproducibility typical of similar investigations of FGK stars. Attention should be given to details of the line formation physics as well as input atomic and molecular data. Collecting high-quality, wide wavelength coverage spectra of M~dwarfs with known fundamental parameters is an essential step of benchmarking spectroscopic parameter determination of low-mass stars.}


   \maketitle
%

\section{Introduction}

M~dwarfs have become a popular subject in the search for exoplanets and they are targets for many current and upcoming missions from the ground as well as from space, such as CARMENES, TESS, PLATO, CRIRES+, SPIRou, and HARPS \citep{2010SPIE.7735E..13Q,TESS2015,2014ExA....38..249R,2016SPIE.9908E..0ID,2020MNRAS.498.5684D,2003Msngr.114...20M}. This interest is due to the small mass and radius of M~dwarfs and their low luminosity, which makes it easier to find planets around them using transit and radial velocity methods. In addition the likelihood of finding a planet in the habitable zone is larger around an M~dwarf since the habitable zone is located closer to the star. It is estimated that each M~dwarf has over 2 planets with a radius between 1-4~R$_\oplus$ \citep{Dressing2015_MdwarfsPlanets}. Over 650 planets have been found around M~dwarfs so far\footnote{Number obtained from NASA exoplanet archive filtering by $\teff$ between 2300 and 3900~$\kelvin$ and stellar radius between 0.10 and 0.56~$\mathrm R_\odot$, September 2020. \\ \url{https://exoplanetarchive.ipac.caltech.edu/}}. M~dwarfs also constitute over 70~\%  of the stars in the solar neighbourhood \citep{Henry2006NrMdwarfs} and are important for the study of the evolution of the elements in the Galaxy due to the long lifespan of M~dwarfs. 

In order to determine the habitability of planets around M~dwarfs, the parameters and abundances of the host star must be accurately determined. Additionally the abundances in the photosphere of a main-sequence star are believed to be a good indicator of the material from which both star and planets were formed \citep[e.g.][]{2015A&A...580A..30T,2017A&A...597A..37D}. 
Accurately determining atmospheric parameters and abundances will advance the theories on both planet formation and galactic evolution.

However, the spectroscopic analysis of M~dwarfs is challenging since the low temperatures cause the optical spectra to be riddled with molecular lines. Lines from titanium oxide increase in strength until mid-type M-dwarfs, and vanadium oxide bands are stronger in later types. Molecular lines from water are found at near-infrared wavelengths for mid- to later types \citep{Grey2009StellarSpectralClass}. In general there are fewer molecular lines in the near-infrared and we can find isolated unblended atomic lines of several elements. However, these wavelength regions are contaminated by telluric lines when observing from the ground. The multitude of molecular lines also makes it difficult to determine the level of the continuum flux needed for spectral analyses. Most M~dwarfs are also highly convective. It is estimated that between 0.3 M$_\odot$ and 0.4 M$_\odot$ they become fully convective, which facilitates generation of strong magnetic fields \citep{2008ApJ...676.1262B,2015ApJ...813L..31Y}. Magnetically sensitive lines are split due to the Zeeman effect which causes a broadening and intensification of the lines. The severity of the split is proportional to the magnetic field strength and the so called Land\'e factor and M~dwarfs have a stronger magnetic field than Sun-like stars by two to three orders of magnitude \citep{Reiners2012MagField}. The faintness is another problem requiring long exposure times for spectroscopic observations. 

In recent years there has been significant progress in spectroscopic studies of M~dwarfs. Using high-resolution near-infrared spectra obtained with modern spectrographs together with techniques for removing telluric lines has made abundance analysis possible. By avoiding magnetically sensitive lines we can accurately determine stellar atmospheric parameters and abundances of individual elements. Recent studies include
\citet[][hereafter L2016]{Lind2016}, \citet[][hereafter L2017]{Lind2017}, 
\citet{2017ApJ...851...26V}, 
\citet[][hereafter P2018]{Pass2018},  \citet[][hereafter P2019]{Pass2019}, 
\citet{Kuznetsov2019ApJ}, 
\citet[][hereafter R2018]{Rajpurohit2018CARMENES}, 
\citet{2019ApJ...879..105L}, 
\citet{2020ApJ...890..133S}, 
and \citet{2020ApJ...892...31B}. 
The parameters derived in these studies do not always agree calling for detailed investigations of the different assumptions and methods used in these studies.

Other observational techniques can constrain the stellar parameters and give a model-independent reference value. Interferometry is one such important technique and with modern interferometers \citet{Boyajian2012} and \citet{Rabus2019} have determined angular diameters for a small number of M~dwarfs. With a known radius and a bolometric flux obtained with photometry and distance from astrometry the Stefan-Boltzmann law is used to determine the effective temperature ($\teff$), and mass-luminosity relations are used to determine the surface gravity ($\log g$). However, the small angular sizes of M~dwarfs present a challenge to the capabilities of modern interferometers, making it difficult to determine precise angular diameters of M~dwarfs and to obtain $\teff$ independently from spectroscopy. Asteroseismology is another valuable technique to obtain stellar parameters such as mass and radius but much is unknown about pulsations in M~dwarfs making it difficult to use this method. \citet{Rodriguez-Lopez2019PulsMdwarf} give a review of the work on observing pulsations in M~dwarfs.

There is a need for benchmark M~dwarfs which can be used for calibrations (sometimes called calibration stars). \citet{Pancino2017} present stars that were chosen as benchmark or calibrator stars for the Gaia-ESO survey (\citealt{2012Msngr.147...25G,2013Msngr.154...47R}). 
These included six M~dwarfs (GJ~205, GJ~436, GJ~526, GJ~551, GJ~581, GJ~699, and GJ~880), three of which are part of the sample being discussed later in this paper. According to \citet{Pancino2017} the benchmark stars should have known parallaxes which today can be easily obtained from the Gaia catalogue \citep{2016A&A...595A...1G,2018A&A...616A...1G}, angular diameters, bolometric fluxes, and homogeneously determined masses. From these properties effective temperature and surface gravity can be determined independently from spectroscopy and then be used to test the spectroscopic methods.

One possible reason for the discrepancies between derived parameters mentioned above
could be departure from local thermodynamic equilibrium, non-LTE effects (NLTE). In this regard, apart from the effects on iron (the usual proxy for overall metallicity), it is interesting to consider departures in \ion{K}{I} lines. This species is known to suffer from strong NLTE effects in optical lines in FGK-type stars \citep{K_NLTE_Reggiani2019}. However, there is less literature available on the effects on infrared lines in M~dwarfs. There is in general a lack of investigations into NLTE effects for M~dwarfs. Of the studies compared in this paper none used NLTE calculations. L2016 and L2017 avoided potassium lines because these showed inconsistencies during the spectroscopic analysis. P2019 state that their model does not fit the core of the observed \ion{K}{I} lines in the near-infrared properly. R2018 give no information regarding NLTE or the fit of potassium lines. The other studies are not discussed in this paper. A possible explanation of the bad fit of \ion{K}{I} lines can be NLTE effects. 

Despite the challenges some studies like \citet{Neves2014MetallicityMdwarfs} and P2018 determined parameters using spectra obtained in the optical. Others used optical combined with near-infrared, such as P2019, and R2018. In this paper we compare the derived parameters between high-resolution spectroscopic studies focusing on the near-infrared using similar methods. In the case of $\teff$ we compare with other methods such as interferometry. In this way the model dependence can be explored and the spectroscopic methods can be improved. We focus on the studies by L2016, L2017, P2018, P2019, and R2018. We also explore possible reasons for discrepancies. L2016, L2017, P2018, and P2019 used similar methods and determined $\log g$ separately from fitting to break degeneracies between metallicity and surface gravity while R2018 included $\log g$ in the fitting. 
For a comparison between high- and low-resolution spectroscopic studies we refer to the above-mentioned works, which compared their results to those of \citet{Mann15} and \citet{Rojas-Ayala2012}, among others.
In \sect{paramcompare} we discuss the studies and compare their parameters. In \sect{analysis} we present a study of non-LTE effects in M~dwarfs. In \sect{sec:discussion} we look into how the different parameters affect synthetic spectra and discuss to what extent non-LTE effects can explain the discrepancies in parameters. We also speculate on other possible reasons for the discrepancies. We end with our conclusions in \sect{conclusion}.


\section{Assessment of previous high-resolution studies of M~dwarfs}
\label{paramcompare}
\subsection{Observations and analysis methods}\label{Sample}
The sample of M~dwarfs that is compared in this paper are the stars from L2016 and L2017, which are also both in P2018, P2019, and R2018. It comprises eleven stars covering a range from early to mid M~dwarfs. The stars and their spectral type can be found in Table~\ref{sampleSNR}. 

The parameters from L2016 and L2017 were derived using observed spectra in the $J$ band obtained with the original CRIRES spectrograph at ESO-VLT with a resolving power of $R \sim 50\,000$.
The signal-to-noise ratio (SNR) of the spectra is given in Table~\ref{sampleSNR}. The parameters from P2018 and P2019 were derived using observed spectra obtained by the CARMENES instrument mounted at the Zeiss 3.5m telescope at Calar Alto Observatory. The CARMENES instrument consists of two spectrographs covering the visible ($520$~nm to $960$~nm) and near-infrared ($960$~nm to $1710$~nm) wavelength range with a spectral resolution of $R\sim 94\,000$ and $R\sim 80\,500$, respectively. In P2018 only the visual wavelength range was used while in P2019 both visual and near-infrared were used separately as well as combined. These spectra have a SNR larger than 75. R2018 used both optical and near-infrared publicly available spectra from CARMENES. These spectra consist of a single exposure \citep{2018A&A...612A..49R} 
and thus have a lower SNR than the co-added spectra used by P2018 and P2019.

   \begin{table}[t]
   \caption{Stars in the sample with signal-to-noise ratio from L2016 and L2017, as well as spectral type from Simbad \citep{Simbad2000}.}
       \label{sampleSNR}
       \centering
       
       \begin{tabular}{l c l }
       \noalign{\smallskip}
        \hline
        \hline
        \noalign{\smallskip}
        Star & SNR & Sp. type \\
        \noalign{\smallskip}
        \hline
        \noalign{\smallskip}
        1 GJ176   & 70    & M2.5 \\
        2 GJ179   & 150   & M2 \\
        3 GJ203   & 65   & M3.5\\
        4 GJ436   & 140  & M3\\
        5 GJ514   & 90  & M1.0\\
        6 GJ581   & 140  & M3\\
        7 GJ628   & 130  & M3\\
        8 GJ849   & 90,120  &M3.5 \\
        9 GJ876   & 100  & M3.5\\
        10 GJ880  & 175  & M1.5\\
        11 GJ908  & 125 & M1 \\
        \noalign{\smallskip}
        \hline
           
       \end{tabular}
   \end{table}
   
   \begin{figure*}
   \centering
   \includegraphics[width=6cm]{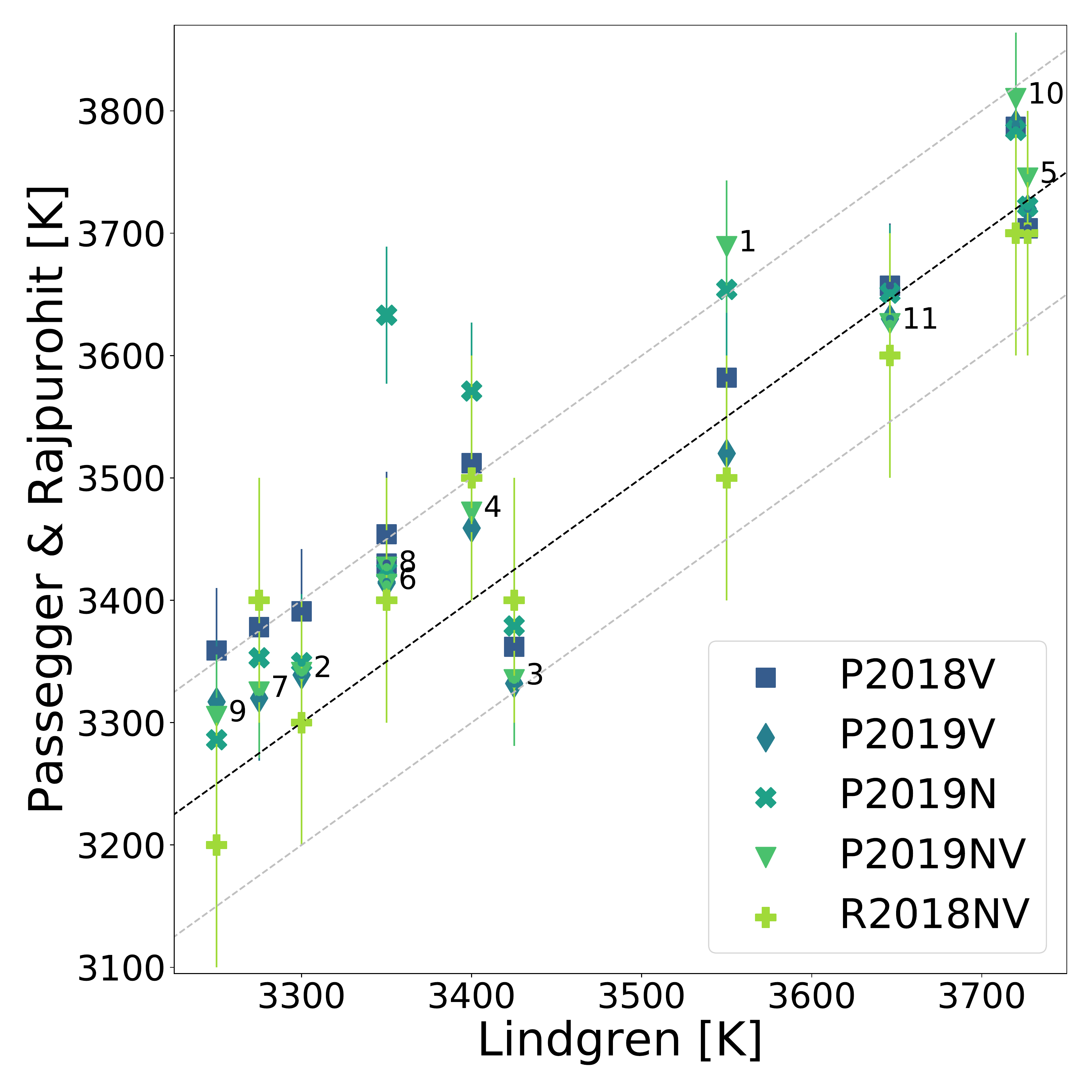}
   \includegraphics[width=6cm]{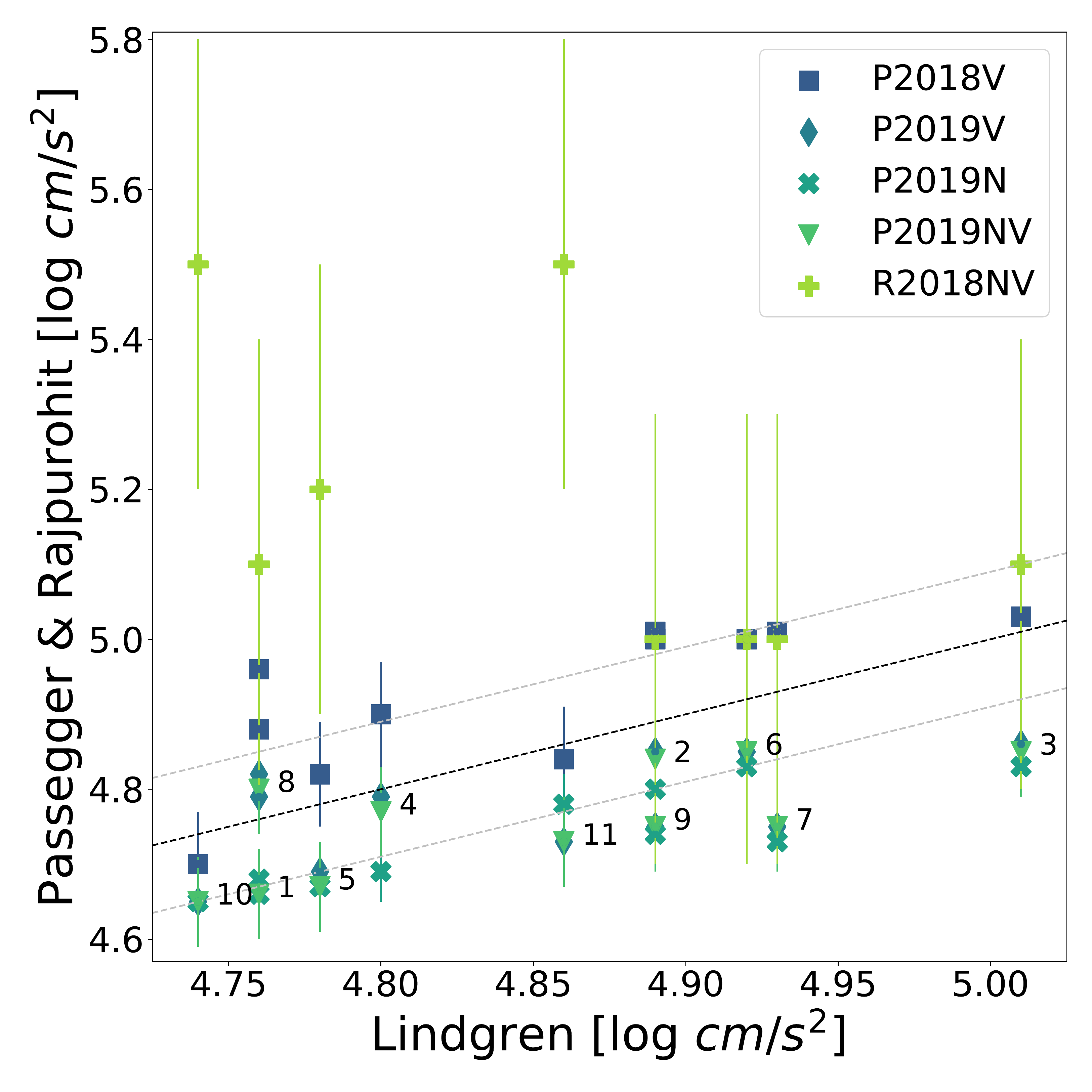}
   \includegraphics[width=6cm]{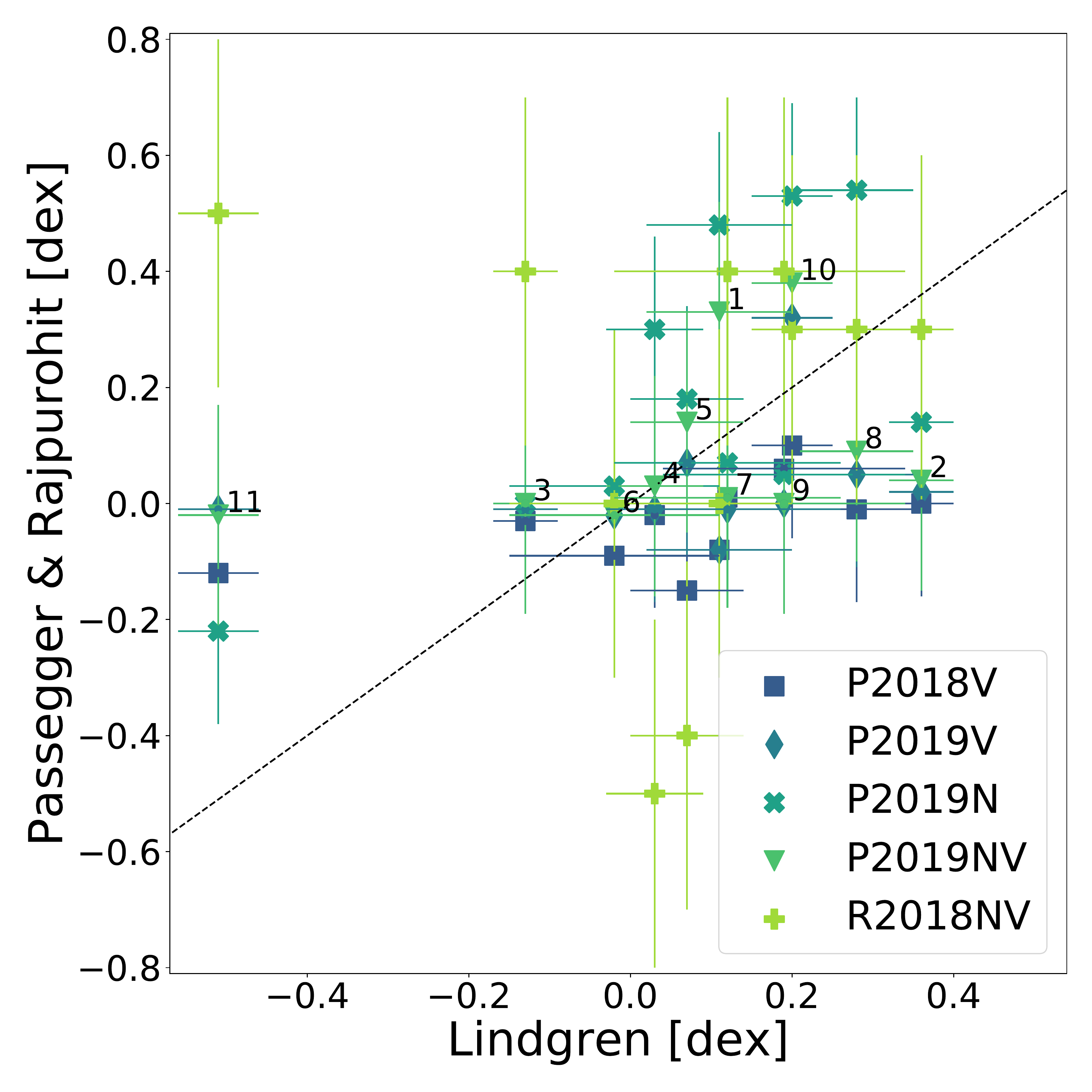}
   \caption{Effective temperature (left), surface gravity (middle), and metallicity (right) derived for stars in common in \citet{Lind2016} and \citet{Lind2017} on the x-axis and \cite{Pass2018,Pass2019, Rajpurohit2018CARMENES} on the y-axis. See Table~\ref{Param} for star numbers and parameter values. The different symbols, designated V, N, and NV in the legend, indicate the wavelength regions (visual, near-infrared, and both combined) which P2018, P2019 and R2018 used to derive the parameters. Error bars indicate corresponding uncertainties. The gray dashed lines indicate the mean uncertainties from L2016 and L2017 for the temperature and gravity. The black dashed diagonal line indicates a 1:1 relation. In the middle plot showing surface gravity the R2018 value for star number 4 is outside of the range of the plot. It has a surface gravity of 4 log $cm/s^2$.}
              \label{PlotParam}%
    \end{figure*}

L2016 and L2017 used the software package Spectroscopy Made Easy, \sme{}, \citep{sme1_1996,sme2_2017} with \marcs{}  \citep{Gustafsson2008A&A} atmospheric models included in \sme{} to derive the stellar parameters. \sme{} computes synthetic spectra on the fly based on a grid of model atmospheres and a line list containing atomic and molecular data. The parameters that are searched for are set as free parameters and $\chi^2$ minimisation is used between the synthetic spectra and the observed spectra to find the best fit. The observations from CRIRES only cover a small wavelength range and there were too few atomic lines for L2016 and L2017 to simultaneously determine effective temperature, surface gravity, and metallicity. The effective temperature and surface gravity were therefore determined prior to the metallicity. L2016 and L2017 determined the effective temperature using fitting of FeH lines. However, this method was only applicable to mid M~dwarfs, as for warmer, earlier types a degeneracy was found between temperature and metallicity. Because of this degeneracy L2017 used temperatures from \citet{Mann15} for the warmest stars (GJ~514, GJ~880, and GJ~908). For the surface gravity L2016 used a $\log g$--mass relation from \citet{Bean2006ApJ} and L2017 used a mass--luminosity relation from \citet{Benedict2016AJ} and radii from empirical relations from \citet{Mann15}. For the metallicity determination L2016 and L2017 fitted lines of Fe, Ti, Mg, Ca, Si, Cr, Co, and Mn in \sme{}, with metallicity and macroturbulence as free parameters. The projected equatorial rotation velocity $\varv\ \sin i$ was set to previous determinations from the literature or to a default value of 1~kms$^{-1}$. The microturbulence was set to a default value of 1~kms$^{-1}$ in L2016, while fixed values based on predictions of published 3D radiation hydrodynamics calculations were used in L2017.

In P2018 and P2019 grids of synthetic spectra based on \phoenix{} atmospheric models were used (ACES and SESAM, see Sect.~\ref{sect:models}). Because of a degeneracy between $T_{\rm eff}$, $\log g$, and [Fe/H], $\log g$ was determined using evolutionary models and $T_{\rm eff}$ -- $\log g$ relations. In P2018 \citet{Baraffe1998A&A} was used and in P2019 they used the PARSEC v1.2S library \citep{2012MNRAS.427..127B,2014MNRAS.444.2525C}. The $\varv\ \sin i$ values were obtained from \citet{2018A&A...614A..76J} 
in P2018 and from \citet{2018A&A...612A..49R} 
in P2019. For the microturbulence ($\varv_{\rm mic}$) P2018 and P2019 used the relation $\varv_{\rm mic}=0.5\cdot \varv_{\rm conv}$, where $\varv_{\rm conv}$ is the convective velocity (which also can be seen as the macroturbulence) obtained from the atmospheric models \citep{2016A&A...587A..19P}. For the fitting procedure P2018 and P2019 used used the $\gamma$-TiO bandhead and lines of K, Ti, Fe, and Mg, as well as Ca (P2019 only). For more information about the methods the reader is directed to P2018 and P2019.

R2018 fitted the whole wavelength region with synthetic spectra using BT-Settl model atmospheres based on the PHOENIX radiative transfer code \citep{2013MSAIS..24..128A}. $T_{\rm eff}$, $\log g$, and [M/H] were all fitted simultaneously. The fitted lines were Ti, Fe, Al, Ca II, K, Na, Mg, and OH.

\subsection{Inferred stellar parameters}
\label{sect:StellarParam}

The stellar atmospheric parameters derived by L2016, L2017, P2018, P2019, and R2018 can be found in Table~\ref{Param}. We show the different parameters in Fig.~\ref{PlotParam}, where the stars are numbered as in Tables~\ref{sampleSNR} and \ref{Param}. The results obtained for different wavelength regions (visual and/or near-infrared) by P2019 are shown separately. The left panel shows the effective temperatures. Most of the measurements agree within uncertainties, although there are some outliers. GJ~849 (number~8) is one such star for which P2019 determined an effective temperature of 3633~K in the near-infrared, which is significantly higher than all others. The derived temperature from L2016 is lower (3350~K) than the others which have temperatures above 3400~K. For all stars in the sample the general trend is that the effective temperatures from P2018 and P2019 are higher than those for L2016, L2017 and R2018, except for GJ~203 (star number~3). We can also see that for the cooler stars P2018V has among the highest temperatures. Note that R2018 has derived the same temperature for star number six and eight. 

The middle panel shows $\log g$, and here we see a larger spread. The $\log g$ values derived in P2018 are often higher than those derived in L2016 and L2017, while the $\log g$ derived in P2019 tends to be lower. The method of determining $\log g$ was improved between P2018 and P2019. The $\log g$ values from P2019V are closer to L2016 and L2017 than those from P2019N and P2019NV. The general trend is that the surface gravities from R2018 are higher than what were derived in the other studies for almost all stars. One exception is star number 4 (GJ~436) for which R2018 gives a surface gravity of 4 log $cm s^{-2}$, note that this is outside the plot. For most of the cooler stars in the sample the surface gravity from R2018 appears to be within the uncertainty in comparison with the other studies (star number 2, 3, 6, 7, and 9). Important to note is that R2018 determined $\rm T_{eff}$, $\log g$, and [M/H] simultaneously. There might therefore be some degeneracy between the parameters.

The right panel of Fig.~\ref{PlotParam} shows the metallicity. Also here we see a larger spread than for the effective temperature, and all values from P2018 and P2019 are more grouped around solar metallicity. Figs.~5 to 7 in P2019 show a similar trend of P2019 metallicities to be shifted towards higher metallicities than the literature values they are comparing against. Metallicities from R2018 are spread over the whole parameter range of the plot and the stars have in many cases a metallicity outside of given uncertainties compared to L2016, L2017, and in fewer cases to P2018 and P2019. R2018 have higher metallicities than P2019 in most cases (excluding P2019N). Two clear exception are star number 4 and 5 (GJ~436 and GJ~514) which have significantly lower metallicites. GJ~436 also shows a discrepancy with the surface gravity. The metallicities by R2018 are more extreme than the other derived metallicities. For many of the stars R2018 have metallicities which magnitudes are outside of 0.3 [dex] while the other studies are inside of this range. Fig.~\ref{PlotParam} also shows that the metallicities determined in the near-infrared by P2019 are generally higher than the other determined metallicites by P2018, P2019, L2016, and L2017, while the P2018 metallicities are among the lowest. There are some outliers: GJ~179 (star number~2), GJ~203 (star number 3), GJ~849 (star number~8), and GJ~908 (star number~11). For GJ~849, P2018 and P2019 has a large spread in metallicity and for two of the values determined by P2018, P2019, and L2016, the difference in metallicities are larger than the quoted uncertainties. R2018 has a metallicity close to what was derived by L2016 for this star. For GJ~203 P2018, P2019, and L2017 agrees while R2018 has a much higher metallicity outside of given uncertainties. This can be connected with degeneracy mentioned above. For GJ~179 and GJ~908 the differences between the metallicities determined by P2018 and P2019 and L2017 are also larger than the quoted uncertainties. The effective temperatures for both of these stars agree within uncertainties for all studies, as do the surface gravity for GJ~179. R2018 obtained a metallicity 1 dex higher for GJ~908 than what was derived by L2017. GJ~908 was discussed as an outlier in P2019 as well. The authors suggested that the reason might be that the star is a member of the thick disk with an age older than that assumed for the evolutionary models that were used to calculate $\log g$. This can not explain the large difference in metallicity since their $T_{\rm eff}$ and $\log g$ agree with the ones determined by L2017. L2017 did not use fitting of FeH lines for this star since it is a warm star and temperatures from \cite{Mann15} were used. \cite{Rojas-Ayala2012} obtained an effective temperature of 3995$\pm$47~K and an overall metallicity of $-$0.41~dex by investigating equivalent widths from low-resolution K-band spectra for the same star. \citet{Mann15} obtained a metallicity of $-$0.45 dex and a $T_{\rm eff}$ of 3646~K using spectrophotometric calibrations. More high-resolution observations of this star are needed to accurately determine its atmospheric parameters. 

The large discrepancy between surface gravity and metallicity that can be seen between R2018 and L2016, L2017, P2018, and P2019 can partly be explained by the difference in method and the quality of the observed data. R2018 used observed spectra from single exposures while P2018 and P2019 used co-added spectra which gives a higher signal-to-noise ratio. L2016, L2017, P2018, and P2019 tried to break degeneracies by determining the surface gravity using empirical calibrations while R2018 determined all three parameters through fitting of synthetic spectra. Investigating the degeneracies between the parameters is outside the scope of this paper and therefore no further detailed comparison is done with R2018 in section \ref{sec:discussion}.

\subsection{Comparison against interferometry} \label{interferometry}
\begin{figure}
   \centering
   \includegraphics[width=9cm]{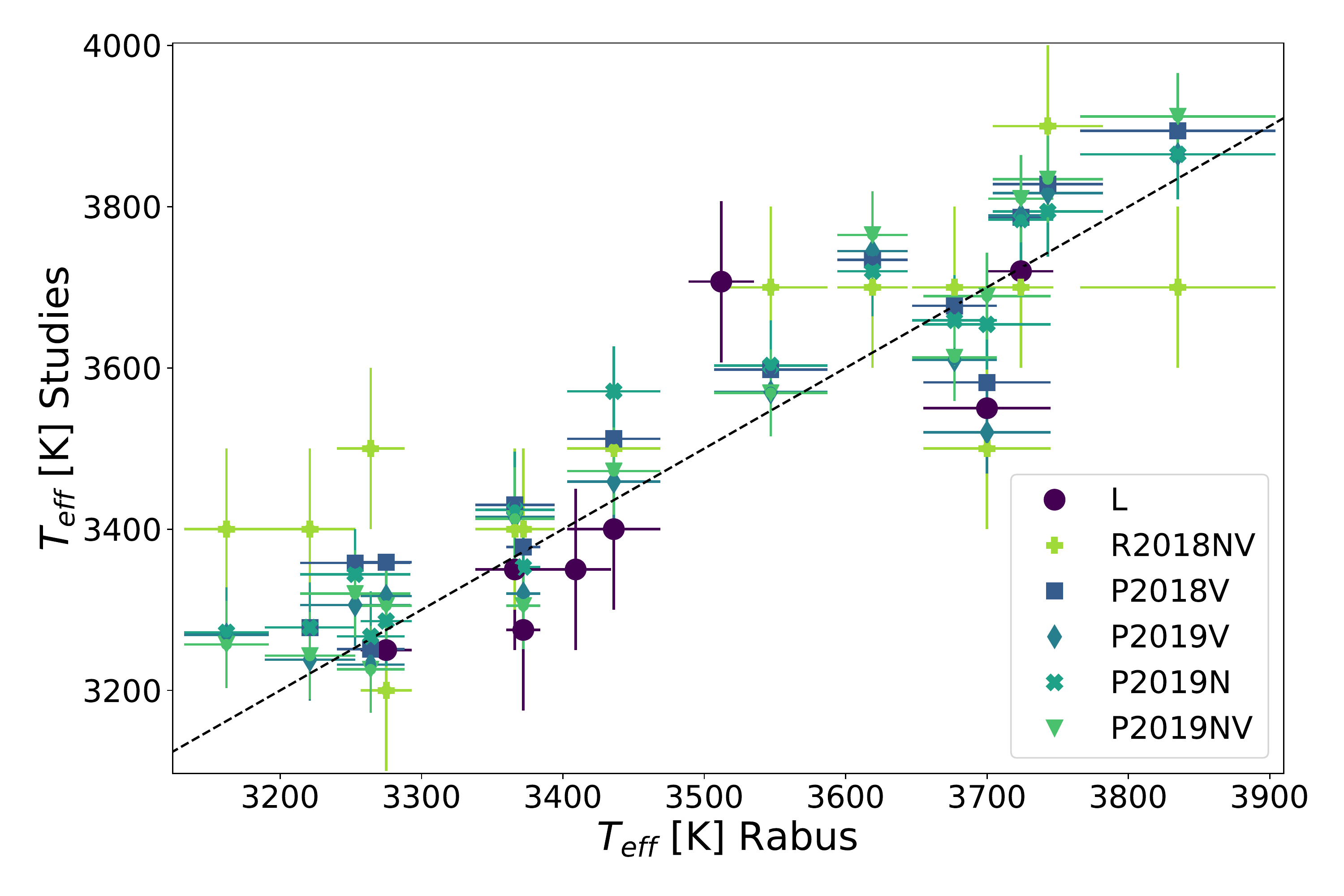}
   \includegraphics[width=9cm]{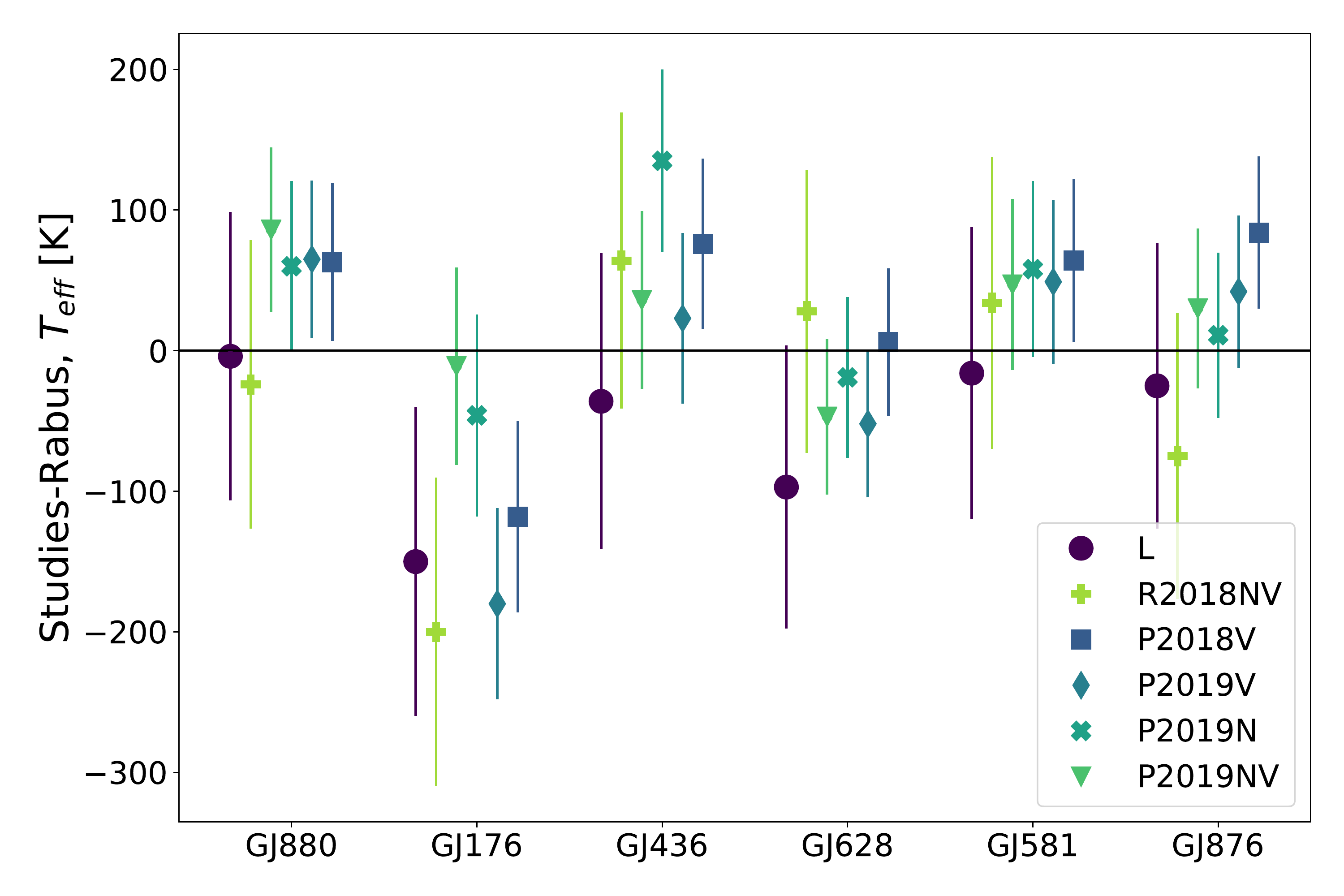}
   
   \caption{Top: Effective temperatures for a sample of stars from Passegger et al. (P2018 and P2019), Rajpurohit et al. (R2018), and/or Lindgren et al. (L2016 or L2017, designated L) that overlaps the sample from \cite{Rabus2019}. Bottom: Difference in effective temperature for the six stars in common between Lindgren et al., Passegger et al., Rajpurohit et al.  and \cite{Rabus2019}. The temperature from \cite{Rabus2019} was subtracted from the others. The error bars indicate the combined uncertainties. V, N, and NV have the same meaning as in Fig.~\ref{PlotParam}.}
              \label{AngTeffDiff}%
    \end{figure}

To further investigate the validity of the derived effective temperature a comparison was done with a method independent from spectroscopy. \cite{Rabus2019} determined the effective temperature using interferometric measurements of the stellar disk from the VLT-Interferometer and photometry from the literature. In \cite{Rabus2019} the effective temperature can be found for six of the eleven stars that are being compared here. These are GJ~176, GJ~436, GJ~581, GJ~628, GJ~876, and GJ~880. The temperatures for these six stars are shown in the upper panel of Fig.~\ref{AngTeffDiff} together with some additional stars found in \citet{Rabus2019} and L2016, L2017, P2018, P2019, and R2018. The differences in effective temperature for the six stars can be seen in the lower panel of Fig.~\ref{AngTeffDiff}.

For all except two stars P2018 and P2019 have a higher temperature than the interferometric determined temperatures, regardless if the temperature was derived using optical or near-infrared wavelengths. The exceptions are GJ~176 and GJ~628 for which the interferometric temperatures are higher. We can see that the $\teff$ values from L2016 and L2017  generally tend to be similar to the interferometric ones although slightly lower. The exceptions are the same two stars as before. R2018 also agrees with \citet{Rabus2019} within uncertainties for the six stars, with one exception which is the same star as mentioned before, GJ~176. In the upper panel of Fig.~\ref{AngTeffDiff} we can see that R2018 has a larger spread in effective temperature than the other studies when comparing to \citet{Rabus2019}.

\citet{Rabus2019} compare their effective temperatures with those derived by \citet{Neves2014MetallicityMdwarfs} who used high-resolution optical spectra and find that the effective temperatures from the optical by \citet{Neves2014MetallicityMdwarfs} are overestimated compared to the near-infrared interferometric effective temperature. It is not clear from Fig.~\ref{AngTeffDiff} that the effective temperatures obtained from spectra in the optical are higher than those obtained in the near-infrared. When looking at the standard deviation of the difference between $\teff$ values for all stars we find that the data coming from the optical have a higher standard deviation than the others, excluding R2018 (70~K for P2018V and 85~K for P2019V versus 52~K for L, 59~K for P2019N and 42~K for P2019NV). R2018 uses the whole wavelength range and has a standard deviation of 89~K. Parameters derived in the optical also have among the highest absolute mean difference (69~K for both P2018V and P2019V versus 55~K for L and P2019N, 43~K for P2019NV). R2018 has a mean difference of 71~K. The combined optical and near-infrared wavelength range from P2019 gave the lowest standard deviation and mean difference while R2018 has the highest. This can indicate that it is better to use near-infrared or combined optical and near-infrared than optical spectra alone.
However, this sample is too small to make a definitive conclusion.
The deviating results of R2018 cannot be taken into account in this regard since they are based on observations of lower quality and a fundamentally different method for determining $\log g$.
More interferometric observations of M~dwarfs are needed so that derived temperatures can be compared to spectroscopically independent temperatures.

\section{Non-LTE effects in M~dwarfs}
\label{analysis}
\label{sect:NLTE}
One aspect that can affect the spectroscopic analysis is departure from local thermodynamic equilibrium (LTE), that is, non-LTE effects (NLTE). This has not yet been investigated in-depth in the case of M~dwarfs. NLTE effects can deepen the cores of some lines for some elements and weaken them for other elements as well as affect the wings. Neglecting NLTE will affect spectroscopically derived parameters or chemical abundances. Recent examples for NLTE studies of potassium including K dwarfs are given by \citet{K_NLTE_Reggiani2019} and \citet{Korotin2020_K_NLTE}. NLTE effects of iron in late-type stars are discussed in \citet{Lind2012_FeNLTE}.

\cite{K_NLTE_Reggiani2019} investigated NLTE effects for three optical and one near-infrared K lines in six stars. The coolest star in the sample was a K dwarf. For all stars in their sample there was a clear difference in line strength between LTE and NLTE for the resonance line at 7699~\AA. Differences were also found for the other lines. However, they were significantly smaller than for the resonance line. For the K dwarf an abundance difference of $-$0.23 dex between LTE and NLTE for the resonance line was determined.

\citet{Lind2012_FeNLTE} investigated NLTE effects of Fe in late-type stars and in their Fig.~2 we can see that for 4000~K, a surface gravity of 4.5, and solar metallicity the difference in abundance for the lines studied by the authors is smaller than 0.01~dex. 

NLTE effects were not included in L2016 and L2017. The synthetic spectra used by P2018 included NLTE for some species for effective temperatures above 4000 K \citep{Husser2013PHOENIX-ACES}, and we assume that NLTE is treated in the same way in the model grid used by P2019, which indicates that the analysis was done in LTE in both cases (see Sect.~\ref{sect:models}). P2019 state that the core of the \ion{K}{I} lines could not be fitted properly. Neither R2018 nor \citet{2013MSAIS..24..128A} give any information regarding NLTE so we assume that their calculations were done in LTE.

In this section we analyse the NLTE effects in M~dwarfs for \ion{K}{I} and \ion{Fe}{I}. Other species will be investigated in later papers.

\subsection{Method}
\label{sect:method}
We used the line formation code Spectroscopy Made Easy (\sme{}; \citealt{sme1_1996,sme2_2017}) version 553 with atmospheric models from the standard \marcs{} grid \citep{Gustafsson2008A&A} and solar abundances from \citet{2007SSRv..130..105G}. Atomic and molecular (OH, MgH, TiO) data were extracted from the \vald{} database\footnote{\url{http://vald.astro.uu.se}} \citep{Pisk:95,2015PhyS...90e4005R}, 
in a wavelength range from $10\,000$ to $16\,000$~\AA\ and around the \ion{K}{I} resonance line at 7699~\AA\ using the default configuration and the ``Extract Stellar'' tool with the following stellar parameters: $T_{\rm eff}$=3000~K, $\log g$=4~[cm\,s$^{-2}$], and solar abundances from \citet{1998SSRv...85..161G} enhanced by +0.5~dex. Data for the most important lines in the wavelength regions investigated below are given in Tables~\ref{tab:lines_7699} and \ref{tab:lines_NIR}.

\sme{} takes departures from LTE into account by interpolating pre-computed grids of departure coefficients. Details of this can be found in Sect.~3 of \citet{sme2_2017}. For potassium, departure coefficient grids were adopted from \citet{2020arXiv200809582A}. These were calculated using the model atom of \citet{K_NLTE_Reggiani2019}, extended down to $\teff=3000\,\kelvin$ and up to $\lgg=5.5\,\dex$. For iron, we adopted grids from \citet{Amarsi2016NLTEFe}, extended down to $\teff=3500\,\kelvin$ and up to $\lgg=5.0\,\dex$. Both go down to $\feh=-5$; the iron grid extends up to $\feh=+0.5$, the potassium grid up to $\feh=+1.0$.

\subsection{Effects on \ion{Fe}{I} lines}
\begin{figure}
   \centering
   \includegraphics[width=4.37cm]{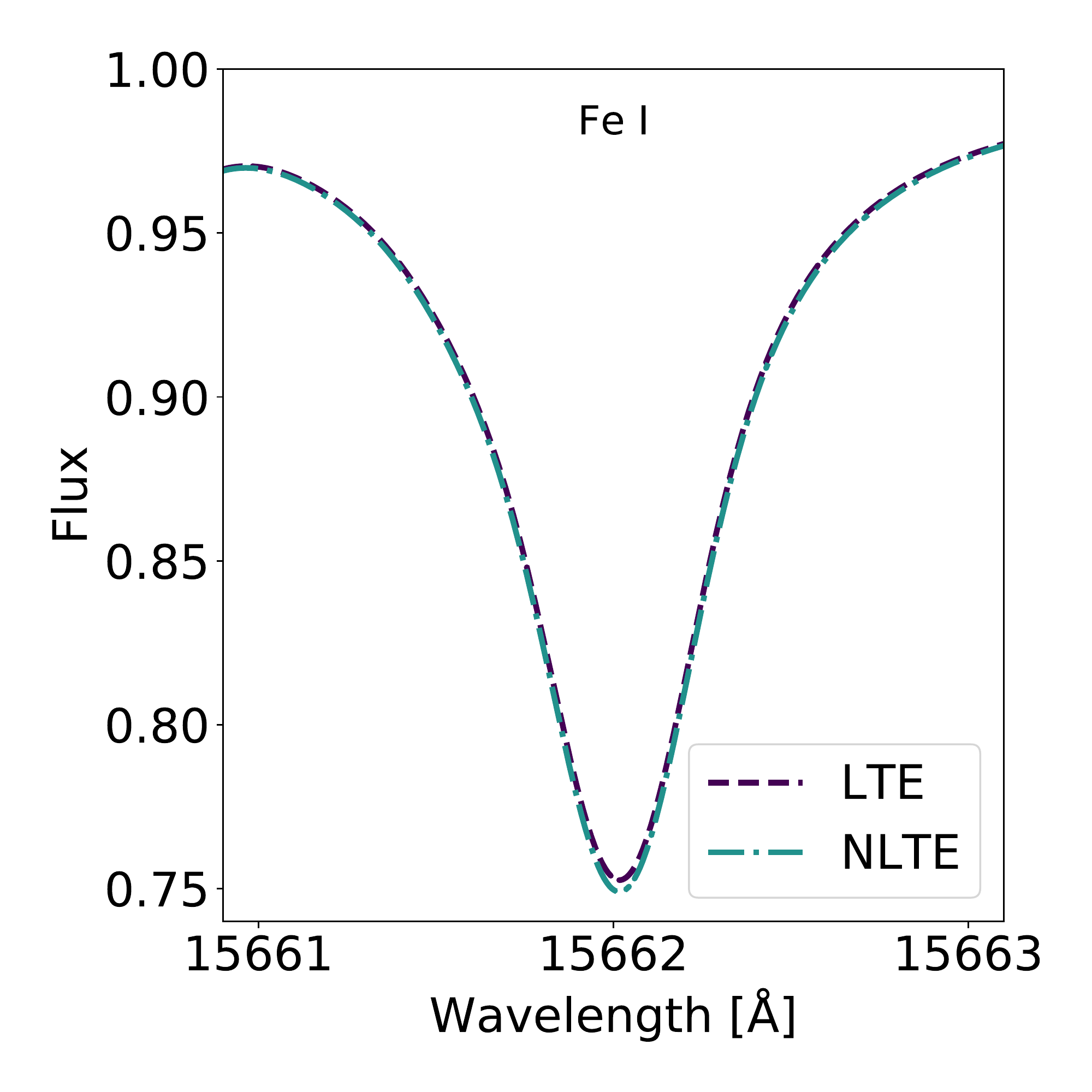}
   \includegraphics[width=4.37cm]{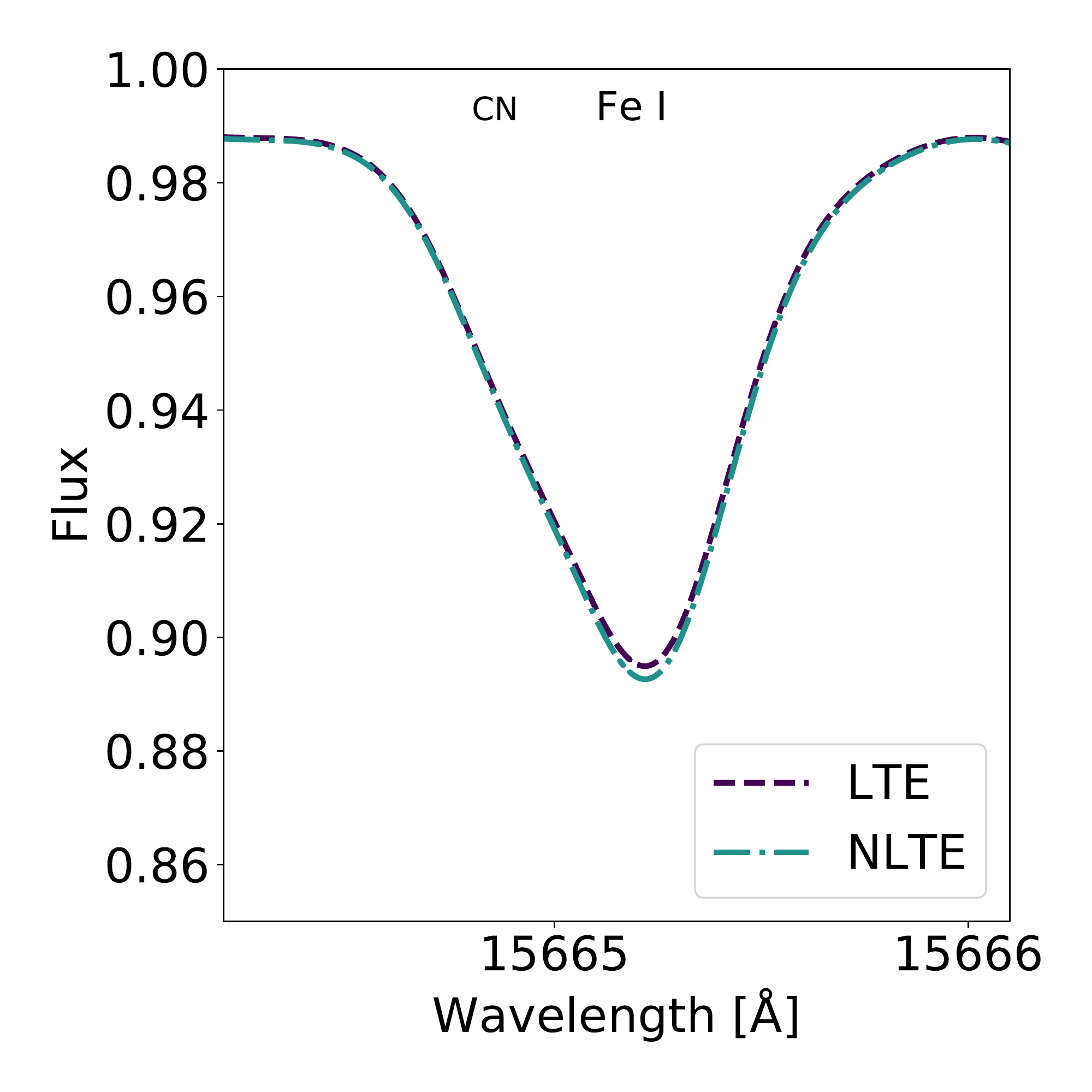}
   
   \caption{Line profiles for two Fe lines in LTE and NLTE. Note the difference in scale on the y-axis. The lines were generated with $T_{\rm eff}$=$3500$~K, $\log g$=4.6, and [M/H]=$0.5$~dex. The weaker line to the right is blended by CN in the blue wing. No observed spectra were available for comparison.}
              \label{FeNLTE}%
    \end{figure}

For iron we generated a small grid of synthetic spectra covering a wavelength range of $10\,000$ to $16\,000$~\AA, with $T_{\rm eff}$ 3500 and 3800~K, $\log g$~4.6 and 4.9 and metallicity 0.5 and $-$0.5~dex, both in LTE and NLTE. We observed the largest difference between LTE and NLTE for the lowest temperature, highest metallicity and lowest gravity. In Fig.~\ref{FeNLTE} we show two example \ion{Fe}{I} line profiles generated for the set of parameters showing the largest difference. These lines are among the lines most affected by NLTE in the near-infrared. As can be seen in the figure, the NLTE line is marginally deeper than the LTE line. The differences in line depth between LTE and NLTE  are 0.48\% for the line at 15662~\AA \  and 0.26\% for the line at 15665~\AA. The difference in equivalent width between LTE and NLTE for the stronger line is 1.7\% and for the weaker line is 2.0\%. The abundance difference between NLTE and LTE for both of these lines is 0.018~dex.  This was found by generating synthetic spectra in \sme{} where the \ion{Fe}{I} abundance was altered so that the equivalent width of the NLTE line matched that of the LTE line. The NLTE abundance had to be lowered since the LTE lines are weaker in the core. The difference for \ion{Fe}{I} between LTE and NLTE is very small for M~dwarfs, which is in accordance with the study by \cite{Lind2012_FeNLTE}. Not in accordance with the study by \cite{Lind2012_FeNLTE} is that we found that the NLTE effect increased with decreasing effective temperatures while they found that it increases with increasing effective temperature. Since the largest difference between LTE and NLTE occurred for lowest temperature it is possible that the NLTE effect in iron will increase when going to even cooler M~dwarfs. A grid of departure coefficients extending to lower temperatures is needed to explore this.


\subsection{Effects on \ion{K}{I} lines}
\label{sec:NLTE_K}

\begin{figure*}
   \centering
   \includegraphics[width=15cm]{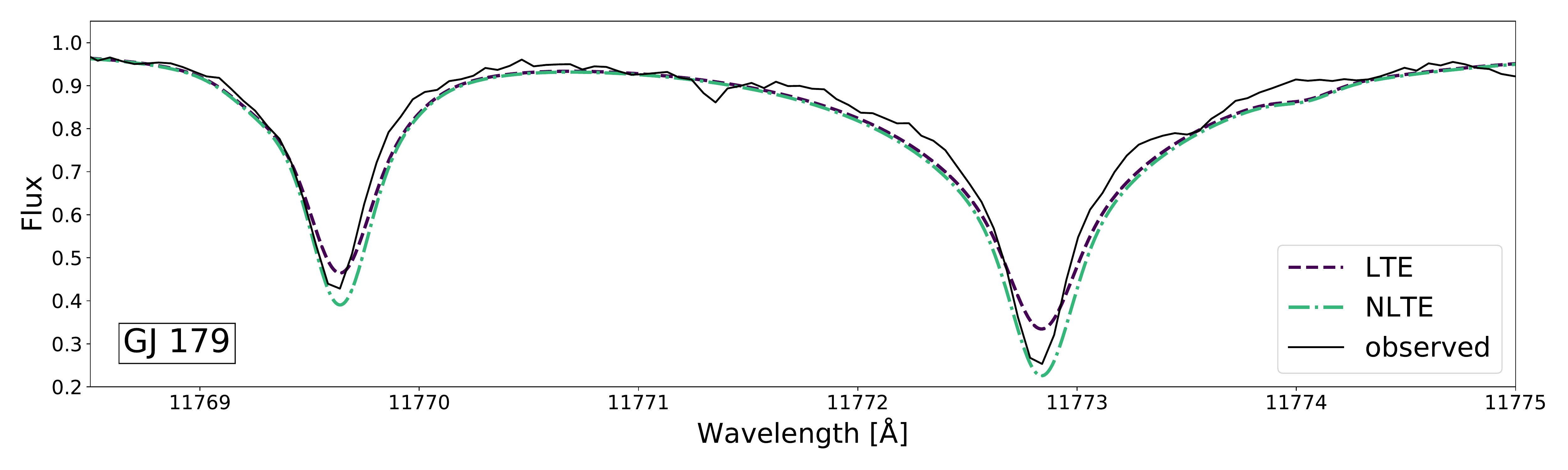}
   \includegraphics[width=15cm]{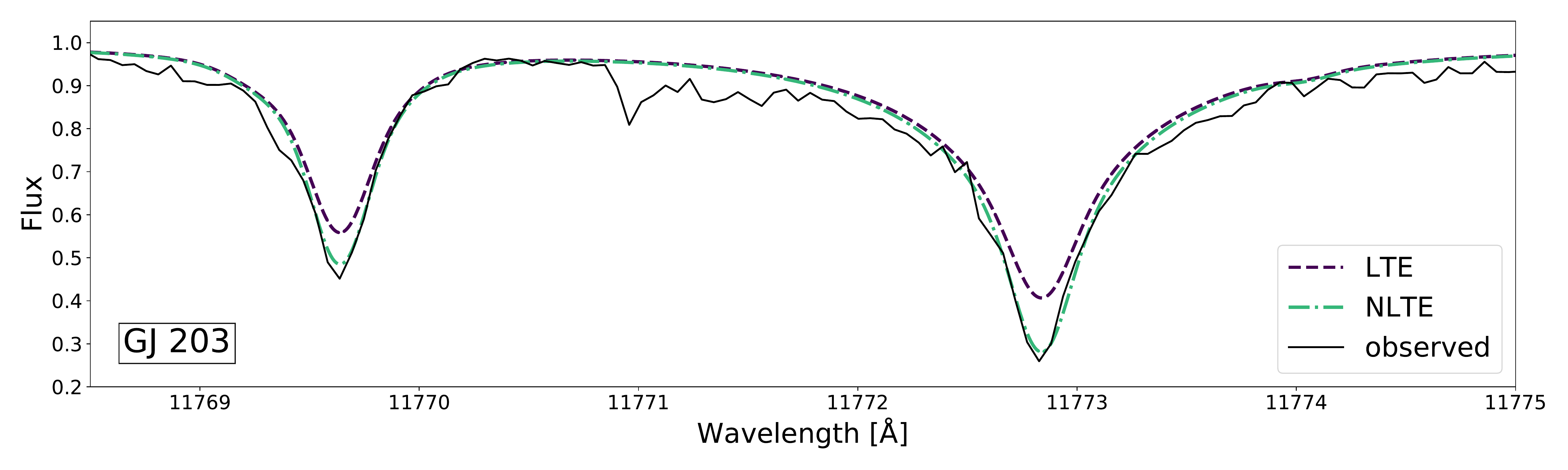}
   \includegraphics[width=15cm]{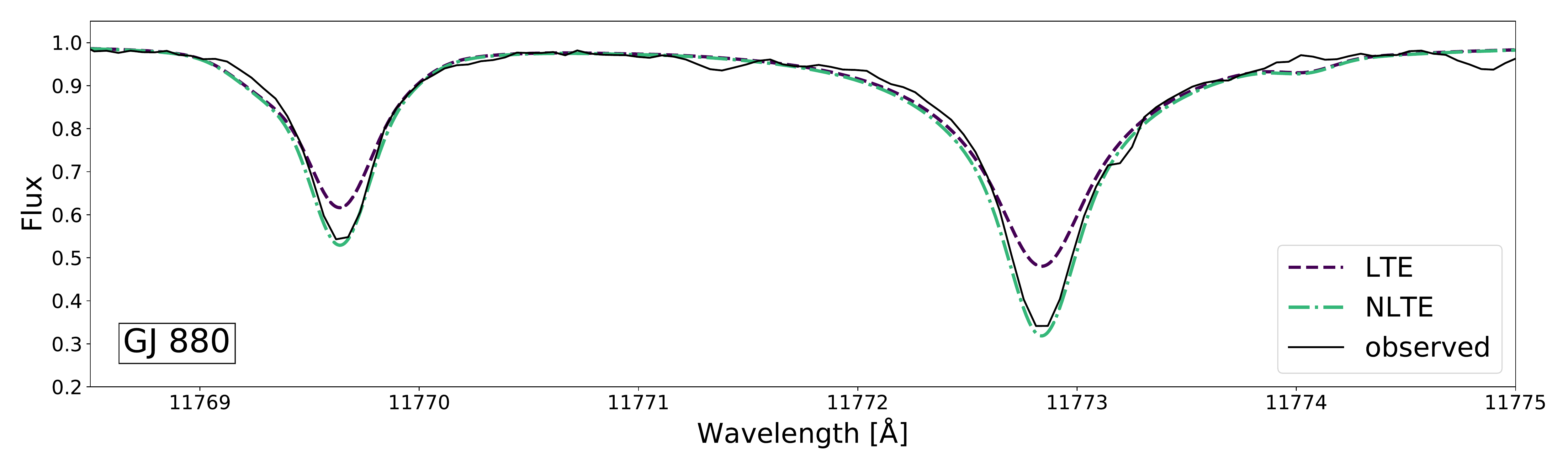}
   \caption{Synthetic spectra showing two K lines generated with parameters for GJ~179, GJ~203, and GJ~880 by L2017 in LTE and NLTE (dashed and dash-dotted lines, respectively). The black solid line shows the observed spectrum for reference. The observed spectrum for GJ~203 has a lower SNR than the other observed spectra.}
              \label{K_LTEvsNLTE}%
    \end{figure*}

Synthetic spectra were generated in the wavelength range $10\,000$ to $16\,000$~\AA \ in LTE and NLTE with parameters from L2017 for the stars GJ~179, GJ~203, GJ~514, GJ~880, and GJ~908 (see Table~\ref{Param}) and then compared. These stars were chosen as they represent both low and high effective temperatures, as well as different metallicities. All of these stars were investigated in L2017 and the observed spectrum used in this comparison is the same as the one used in L2017. The K lines in the near-infrared most affected by NLTE can be seen in Table~\ref{Klines}, which also shows the percentage difference between the equivalent width in LTE and NLTE as well as the reduced equivalent width for the stars with the strongest and weakest lines. In Fig.~\ref{K_LTEvsNLTE} we show the line profiles of two of the K lines in LTE and NLTE for three stars (GJ~179, GJ~203, and GJ~880). As can be seen in the figure, the NLTE lines are significantly deeper than the LTE lines and match the observed lines better. This is in accordance with theory which states that NLTE effects for K are mostly driven by photon losses which generate an overpopulation in the lower energy levels of the transitions, making the lines deeper in NLTE \citep{Asplund2005_NLTE}. The largest differences between LTE and NLTE occur for the strongest lines in the near-infrared which also have the lowest transition energies, see Table~\ref{Klines}. We also included the resonance line at 7699~\AA \ which was examined in previous studies of NLTE in FGK stars  \citep{K_NLTE_Reggiani2019, Korotin2020_K_NLTE}. This line has a smaller difference in equivalent width than the lines in the near-infrared. The line is severely blended with TiO and the NLTE line is saturated. The importance of the NLTE effect on other potassium lines in optical spectra of M~dwarfs is beyond the scope of this work and is left to be determined in future studies. 

   \begin{table*}[t]
    \caption{Difference in equivalent width between LTE and NLTE for K lines in the near-infrared with wavelength $\lambda$. The second column gives the energy of the lower level of the transition ($E_{\rm low}$). The last two columns show the reduced equivalent width calculated from the LTE synthetic spectra for GJ~179 and GJ~908, which have the strongest and the weakest lines, respectively.}
       \label{Klines}
       \centering
       \begin{tabular}{r r r r r r r| r r}
       \hline
       \hline
       \noalign{\smallskip}
       $\lambda$ [\AA]  & $E_{\rm low}$ [eV]  & GJ 179 [\%] & GJ 203 [\%] & GJ 514 [\%] & GJ 880 [\%] & GJ 908 [\%] & GJ 179 eq.w & GJ 908 eq.w\\
       \noalign{\smallskip}
       \hline
       \noalign{\smallskip}
       7699.0  & 0.000 & -1.3 & -2.4 & -3.6  & -2.9  & -4.9 & -3.47 & -3.85\\
       11019.8 & 2.670 & -1.1 & -1.1 & -1.4  & -1.4  & -0.6 & -4.95 & -5.79\\
       11022.6 & 2.670 & -1.2 & -1.1 & -1.4  & -1.6  & -0.8 & -5.09 & -5.93\\
       11690.2 & 1.610 & -4.9 & -6.8 & -8.5  & -8.0  & -7.8 & -4.15 & -4.39\\
       11769.6 & 1.617 & -6.6 & -9.2 & -12.1 & -11.3 & -13.4 & -4.49 & -4.94\\
       11772.8 & 1.617 & -5.9 & -8.9 & -12.9 & -11.7 & -13.7 & -4.11 & -4.50\\
       12432.3 & 1.610 & -6.3 & -9.1 & -12.5 & -11.5 & -14.4 & -4.39 & -4.88\\
       12522.1 & 1.617 & -5.5 & -8.2 & -11.0 & -9.9  & -12.0 & -4.19 & -4.61\\
       13377.8 & 2.670 & -1.6 & -0.2 & -2.2  & -1.2  & ...   & -6.04 & -6.83\\
       15163.1 & 2.670 & -0.8 & -0.6 & -0.9  & -0.9  & -0.6  & -4.39 & -4.87\\
       15168.4 & 2.670 & -0.8 & -0.6 & -1.0  & -0.8  & -0.6  & -4.47 & -5.02\\
       \noalign{\smallskip}
       \hline
       \end{tabular}
    \tablefoot{The differences in equivalent width are calculated as (LTE$-$NLTE)/NLTE$\cdot$100. The \ion{K}{I} line at 13\,377.8 \AA \ in the star GJ~908 was too weak for the measurement of equivalent widths. The reduced equivalent widths are calculated by $\log$(eq.width$_{\rm LTE}$/$\lambda$).}
   \end{table*}
   
   \begin{table*}[t]
    \caption{Abundance corrections (differences in abundance for NLTE$-$LTE in dex) for K lines determined by matching the equivalent widths (columns ``Eq wi'') and fitting the line profiles (columns ``Fit'') of K lines at wavelength $\lambda$. The line at 11\,690.2 $\AA$ is blended with \ion{Fe}{I}.}
       \label{KlinesAbundCorr}
       \centering
       \begin{tabular}{p{0.05\linewidth}c c c c c c c c c c}
       \hline
       \hline
       \noalign{\smallskip}
       $\lambda$ [\AA]  &\multicolumn{2}{c}{GJ 179} & \multicolumn{2}{c}{GJ 203} & \multicolumn{2}{c}{GJ 514} & \multicolumn{2}{c}{GJ 880} & \multicolumn{2}{c}{GJ 908}\\
         &\rm Eq wi & Fit & Eq wi & Fit & Eq wi & Fit & Eq wi & Fit & Eq wi & Fit\\
       \noalign{\smallskip}
       \hline
       \noalign{\smallskip}
       7699.0  &-0.065 & ...   & -0.060  & ...    & -0.074 & ...    &-0.069 & ...    & -0.069 & ... \\
       11019.8 &-0.007 & ...   & -0.005  & ...    & -0.008 & ...    &-0.008 & ...    & -0.004 & ...\\
       11022.6 &-0.007 & ...   & -0.005  & ...    & -0.008 & ...    &-0.009 & ...    & -0.004 & ...\\
       11690.2 &-0.076 & -0.108 & -0.119 & -0.138 & -0.222 & -0.372 &-0.208 & -0.270 & -0.192 & -0.258\\
       11769.6 &-0.085 & -0.127 & -0.104 & -0.130 & -0.157 & -0.211 &-0.157 & -0.195 & -0.119 & -0.141\\
       11772.8 &-0.072 & -0.077 & -0.104 & -0.089 & -0.158 & -0.165 &-0.141 & -0.137 & -0.151 & -0.203\\
       12432.3 &-0.074 & ...   & -0.094  & ...    & -0.134 & ...    &-0.130 & ...    & -0.107 & ...\\
       12522.1 &-0.061 & ...   & -0.089  & ...    & -0.140 & ...    &-0.126 & ...    & -0.118 & ...\\
       13377.8 &-0.006 & ...   & -0.004  & ...    & -0.006 & ...    &-0.009 & ...    & -0.004& ...\\
       15163.1 &-0.009 & ...   & -0.006  & ...    & -0.008 & ...    &-0.009 & ...    & -0.004& ...\\
       15168.4 &-0.008 & ...   & -0.005  & ...    & -0.007 & ...    &-0.008 & ...    & -0.004& ...\\
       \hline
       \end{tabular}
   \end{table*}
   
   \begin{figure}
   \centering
   \includegraphics[width=9cm]{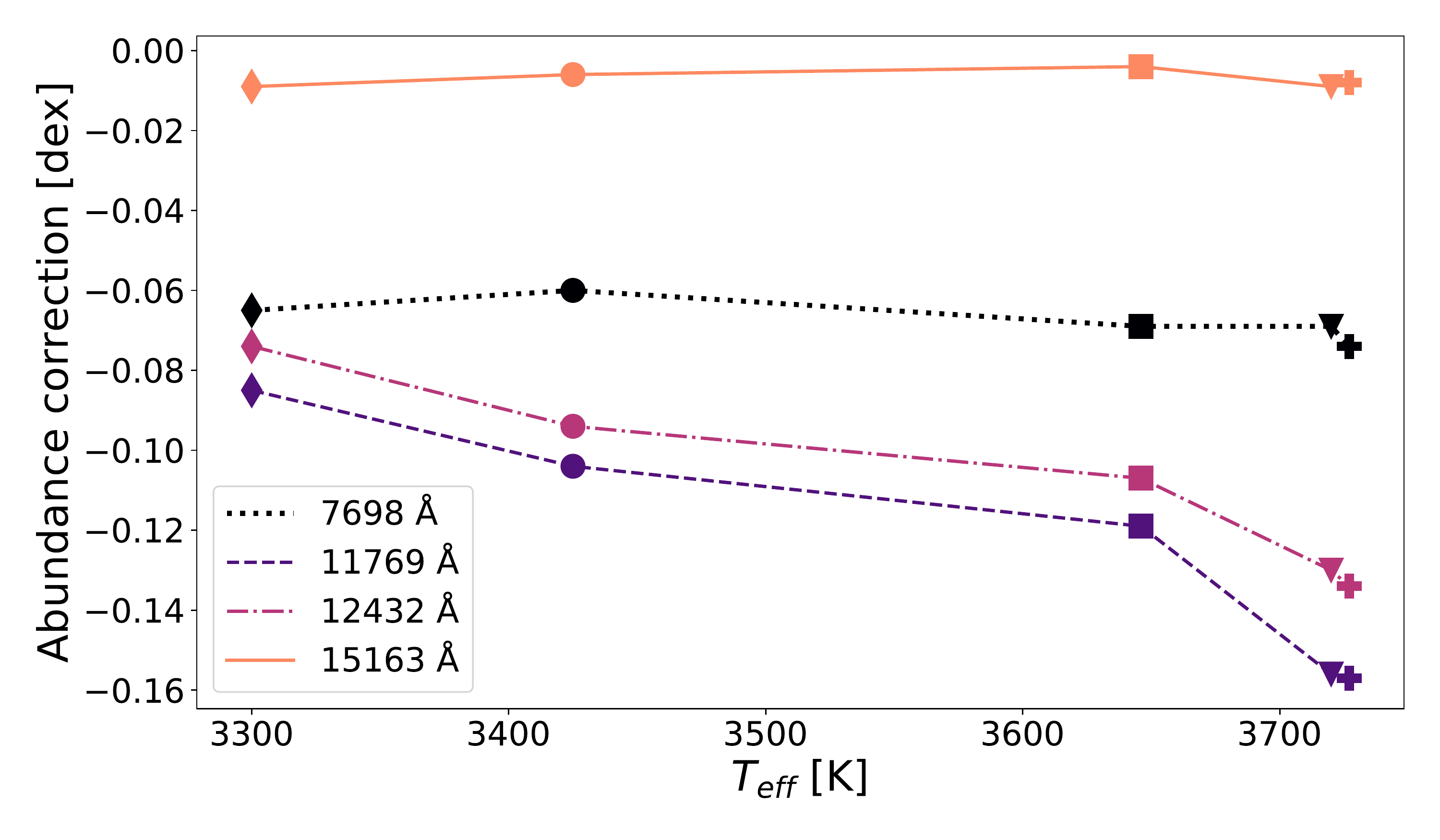}
   \includegraphics[width=9cm]{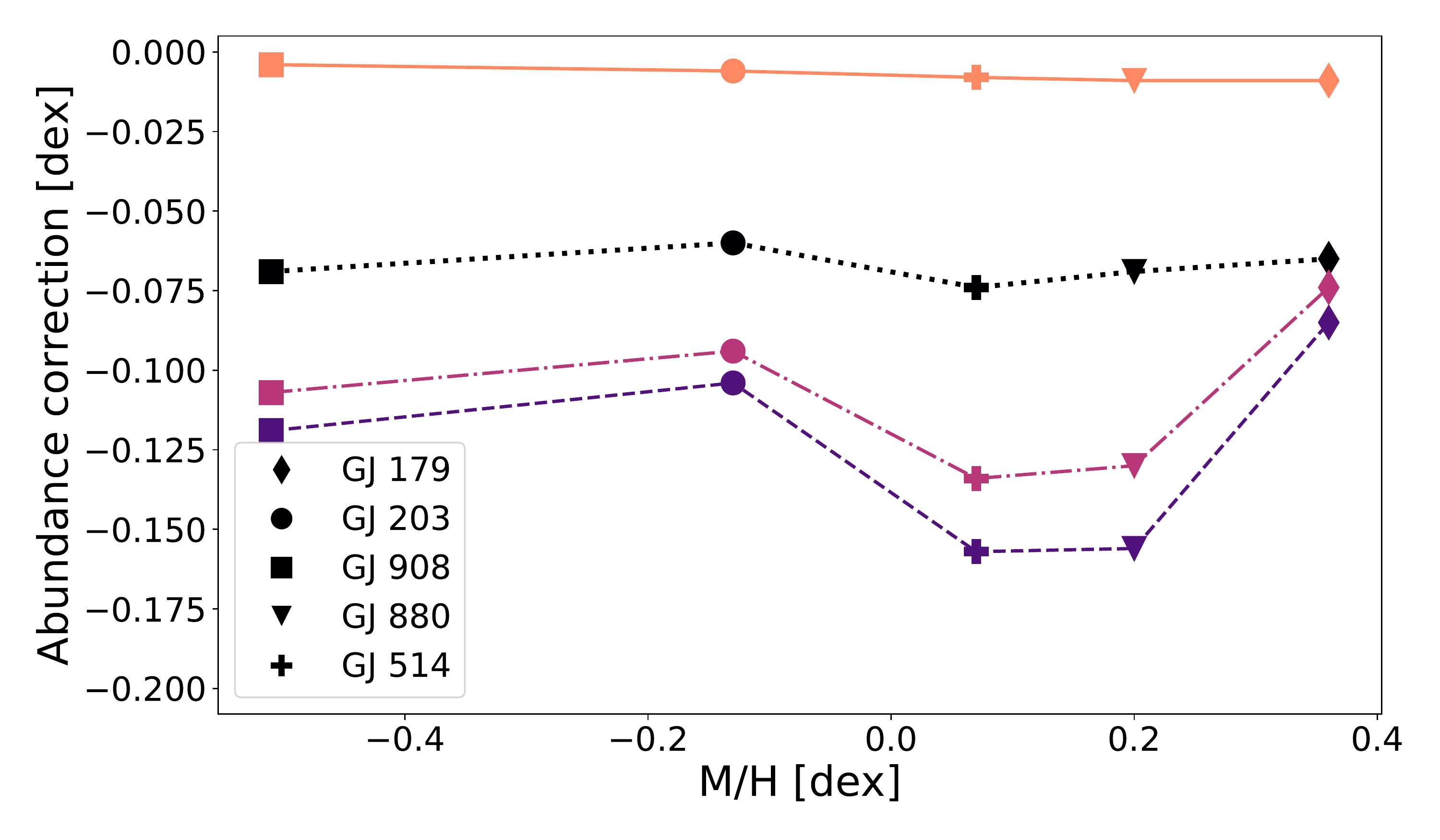}
   \caption{Abundance corrections for four different potassium lines from Table~\ref{KlinesAbundCorr} versus effective temperature and metallicity for the stars in the table. The legend in the top panel shows the line styles used to indicate different spectral lines, and the legend in the lower panel shows the symbols used to indicate different stars.}
              \label{K_AbundCorrTeff}%
\end{figure}

\begin{figure*}
   \centering
   \includegraphics[width=15cm]{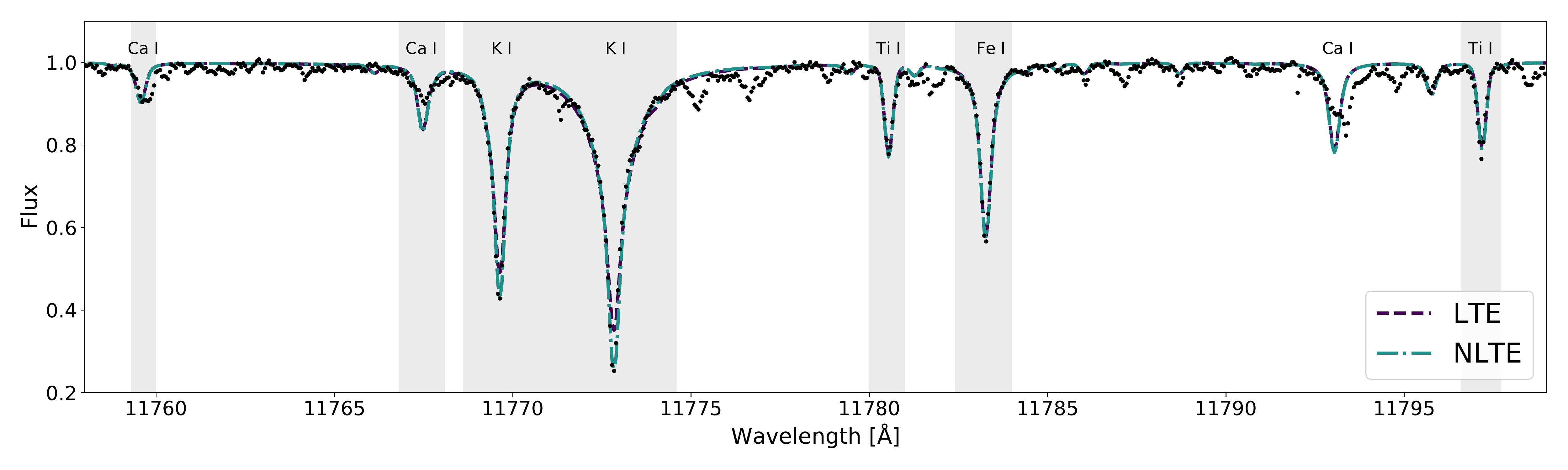}
   \caption{Synthetic spectra for the star GJ~179 for best-fit $T_{\rm eff}$ in LTE and NLTE (dashed and dot-dashed lines respectively). Black points indicate the observed spectrum. Shaded areas show fitted regions.}
              \label{Tefffree}%
    \end{figure*}

To further test the differences between LTE and NLTE an abundance analysis was performed for K for the same five stars as above. Individual K lines in the near-infrared were fitted one at a time by varying the abundance of K in \sme{} using the parameters from L2017 (found in Table~\ref{Param}). 
The fit was done against CRIRES observed spectra in two small wavelength ranges (11\,670--11\,730~\AA \ and 11\,750--11\,800~\AA), one of which is shown in Fig.~\ref{Tefffree}. The observed spectra are the same as used in L2017. For these five stars only three K lines are available in the observed spectra (one is strongly blended with an iron line). The abundance differences obtained in the fitting between NLTE and LTE for these three lines can be seen in Table~\ref{KlinesAbundCorr} (columns ``Fit'').

Since the observed spectra covered few K lines we expanded the investigation to all K lines which showed a clear NLTE effect in the near-infrared (as seen in Table~\ref{Klines}), with the addition of the resonance line at 7699~\AA \ used by P2018 and P2019. This was done by generating an LTE synthetic spectrum using \sme{} with the same settings as in the previous test. We then generated NLTE synthetic spectra and altered the abundance of K to match the equivalent width of the K lines in the LTE synthetic spectrum. We had to lower the NLTE abundances in order to mimic the equivalent widths measured from the LTE spectrum. However, we emphasise that the line shapes of the LTE and NLTE synthetic spectra were significantly different. Therefore the abundance differences should be interpreted with caution. The result can be seen in Table~\ref{KlinesAbundCorr} (columns ``Eq wi''). The abundance correction derived from fitting is higher than the abundance correction obtained using equivalent widths. The reason for this is unknown.  
The three warmest stars (GJ~514, GJ~880, and GJ~908) show the largest difference between NLTE and LTE.
\citet{Korotin2020_K_NLTE} present similar results as in Tables~\ref{Klines} and \ref{KlinesAbundCorr} for lines in the near-infrared for FGK stars. They find that the \ion{K}{I} lines at 11\,769~\AA, 11\,772~\AA, 12\,432~\AA, and 12\,522~\AA \ need to be calculated in NLTE while the lines at 15\,163~\AA \ and 15\,168~\AA \ can be treated in LTE.

We show the abundance corrections derived using equivalent widths for some of the K lines in \fig{K_AbundCorrTeff}. Each colour represents one potassium line and each symbol represents a star. In the upper panel we can clearly see that for the stronger lines at 11\,769~\AA \   and 12\,432~\AA \ the abundance correction increases with the effective temperature. The resonance line at 7699~\AA \ has a similar correction for all temperatures, as does the weaker line at 15\,163~\AA. In the lower panel we can see how the abundance correction changes with metallicity. We see that for the two strong lines the abundance corrections are larger for 0.0 and 0.2~dex than for the other metallicities. The two stars corresponding to these points are the two warmest stars. Thus it seems that the NLTE effects vary more with effective temperature than with metallicity.

\citet{Asplund2005_NLTE} states that for the resonance line the largest abundance correction between LTE and NLTE occurs for the highest effective temperatures and lowest surface gravities. We observe that the two warmest stars (GJ~514 and GJ~880) in Table~\ref{Klines} also have among the lowest surface gravities of the sample. The abundance corrections for these stars are among the highest. 

\subsection{Effects on inferred stellar parameters} \label{NLTEeffectTeffMet}
   \begin{table}[t]
   \caption{Difference in effective temperature and metallicity between NLTE and LTE for five stars (NLTE$-$LTE) derived from fitting synthetic to observed spectra in the wavelength region shown in Fig.~\ref{Tefffree}.}
       \label{Tab:NLTEParamDiff}
       \centering
       
       \begin{tabular}{l c c }
       \noalign{\smallskip}
        \hline
        \hline
        \noalign{\smallskip}
        Star & $\Delta \teff$ [K]& $\Delta$ [M/H] [dex] \\
        \noalign{\smallskip}
        \hline
        \noalign{\smallskip}
        GJ 179   & 88  & -0.24 \\
        GJ 203   & 50  & -0.20\\
        GJ 514   & 153 & -0.16\\
        GJ 880   & 213 & -0.15\\
        GJ 908   & 133 & -0.19\\
        \noalign{\smallskip}
        \hline
           
       \end{tabular}

   \end{table}
In order to estimate how NLTE would affect the spectroscopically derived parameters by L2016, L2017, P2018, P2019, and R2018 we investigated the differences in $\teff$ and metallicity obtained when LTE is used for fitting K lines compared to when NLTE is used. In P2018 one out of fifteen lines was \ion{K}{I} while in P2019 five out of fifteen lines were \ion{K}{I} in the wavelength region classified as near-infrared by the authors. L2016 and L2017 avoided \ion{K}{I} lines. R2018 uses several K lines but not as big proportions as P2019. Since a considerable number of near-infrared potassium lines were used in P2019 it is important to estimate how NLTE would affect the parameters derived in this case. We fitted a synthetic spectrum to an observed spectrum with \sme{} with $\teff$ as a free parameter for the same stars as in Sect.~\ref{sec:NLTE_K} in a short wavelength range including K lines in LTE and NLTE (see Fig.~\ref{Tefffree}). The other parameters were set to values from L2017, see Table~\ref{Param}. Since the NLTE effects of \ion{Fe}{i} were found to be minor, this species was fixed to LTE in all cases.

As can be seen in Fig.~\ref{Tefffree} there is a clear difference between NLTE and LTE for the K lines but the difference is small for the other lines. In Table~\ref{Tab:NLTEParamDiff} we list the differences in $\teff$ and metallicity for the five stars. These differences should only be seen as indications of the possible NLTE effect on the effective temperature and metallicity and not actual temperature and metallicity corrections for these stars. This is because only a short wavelength range was used. We find that LTE underestimates the effective temperature compared to NLTE. For the coolest star (GJ~179) the effective temperature derived in NLTE is 88~K higher than the LTE effective temperature while for one of the warmest stars (GJ~880) the NLTE effective temperature is 213~K higher. For the warmest star (GJ~514) the difference is 153~K. This indicates that the temperatures derived by P2018 and P2019 might be underestimated and that the discrepancy is larger for the warmer stars. This effect could be larger in P2019 since a larger number of \ion{K}{I} lines were used in the fitting in the near-infrared. However, increasing the $\teff$ for both P2018 and P2019 data would increase the difference in temperature between P2018 and P2019 and L2016 and 2017 and between P2018 and P2019 and  \cite{Rabus2019}, not decrease it. For R2018 temperatures the effect would be similar as for P2018 and P2018 but not as severe since they use more lines from other elements in the fitting. Increasing the effective temperature for the stars in the sample would lower the discrepancy for a few stars but increase it for a majority of the stars.

For the metallicity we find that the LTE metallicity is higher than the NLTE metallicity. The largest difference between LTE and NLTE occurs for GJ~179 which is the coolest star of this sample and has the highest metallicity. The second coolest star (GJ~203) has the second largest difference between LTE and NLTE metallicity. This star has a low metallicity and the highest surface gravity in the sample. The two warmest stars (GJ~514 and GJ~880) have among the lowest surface gravities and have the lowest difference in metallicity between LTE and NLTE. It is interesting to note that GJ~880 has a higher metallicity and shows a larger difference in $\teff$ than GJ~514 even though they have a similar effective temperature. Lowering the metallicity in the P2019N (near-infrared) sample would decrease the difference between P2019N and L2016 and L2017 for about half of the stars. Two of these stars were identified as outliers above (GJ~849 and GJ~908). For GJ~849 the metallicity from P2019N is much higher than the others. Lowering this metallicity would make it more in line with the other results from P2018 and P2019, as well as the metallicity from L2016. However, this star also has a discrepancy in the derived effective temperatures (see Sect.~\ref{sect:StellarParam}). For GJ~908 the metallicity from P2019N is closer to L2017 than the other metallicity estimates from P2018 and P2019. Lowering it further would make the metallicity more in line with both L2017, \cite{Rojas-Ayala2012}, and \citet{Mann15}. However, it would increase the difference to the other metallicities from P2018 and P2019. Lowering the metallicity of R2018 would also decrease the difference between R2018 and L2016 and L2017 for about half of the stars. This is especially true for GJ~908 which shows a very large difference between R2018 and L2017. Two strong exceptions are GJ~436 and GJ~514 that were given as outliers in section \ref{paramcompare}.

As mentioned in Sect.~\ref{sect:StellarParam} the metallicities from P2019N are generally higher than the other metallicities and would therefore benefit the most from a downward correction of metallicity. When looking at Figs.~5 and 6 of P2019 we can see that increasing the effective temperatures of P2019 by 100--200~K would worsen the fit to the literature values while lowering the metallicity by 0.15~dex might improve it.

\section{Possible explanations for parameter discrepancies}
\label{sec:discussion}
\subsection{Comparing synthetic and observed spectra}\label{sec:CompSynth}
In order to further investigate the difference in the derived atmospheric parameters we generated synthetic spectra in LTE using \sme{}, \marcs{} cool-stars atmospheric models, and the line list described in Sect.~\ref{sect:method}. In addition, the line data for FeH used by L2017 were added\footnote{Available on the \marcs\ webpage at \url{https://marcs.astro.uu.se/documents.php}.}. Solar abundances from \citet{2007SSRv..130..105G} were used. The parameters given in Table~\ref{Param} were used. Synthetic spectra were not generated with parameters from R2018 because of the differences in methods, see section \ref{paramcompare}. For the synthetic spectra generated with the P2018 and P2019 parameters the microturbulence was estimated from Fig.~3 in \cite{Husser2013PHOENIX-ACES} and for the macroturbulence the relation mentioned in Sect.~\ref{Sample} was used. We used the turbulences given in L2016 and L2017 for the corresponding parameters.  The generated synthetic spectra were then compared to observed spectra obtained by the CRIRES spectrograph at the VLT (the same spectra were used in L2016 and L2017). There are spectra available from CARMENES \citep{2018A&A...612A..49R} with a lower SNR than used in P2018 and P2019 and riddled with telluric lines. For these reasons we do not include a comparison to these spectra in this investigation.

\begin{figure}
   \begin{center}
   \includegraphics[width=9cm]{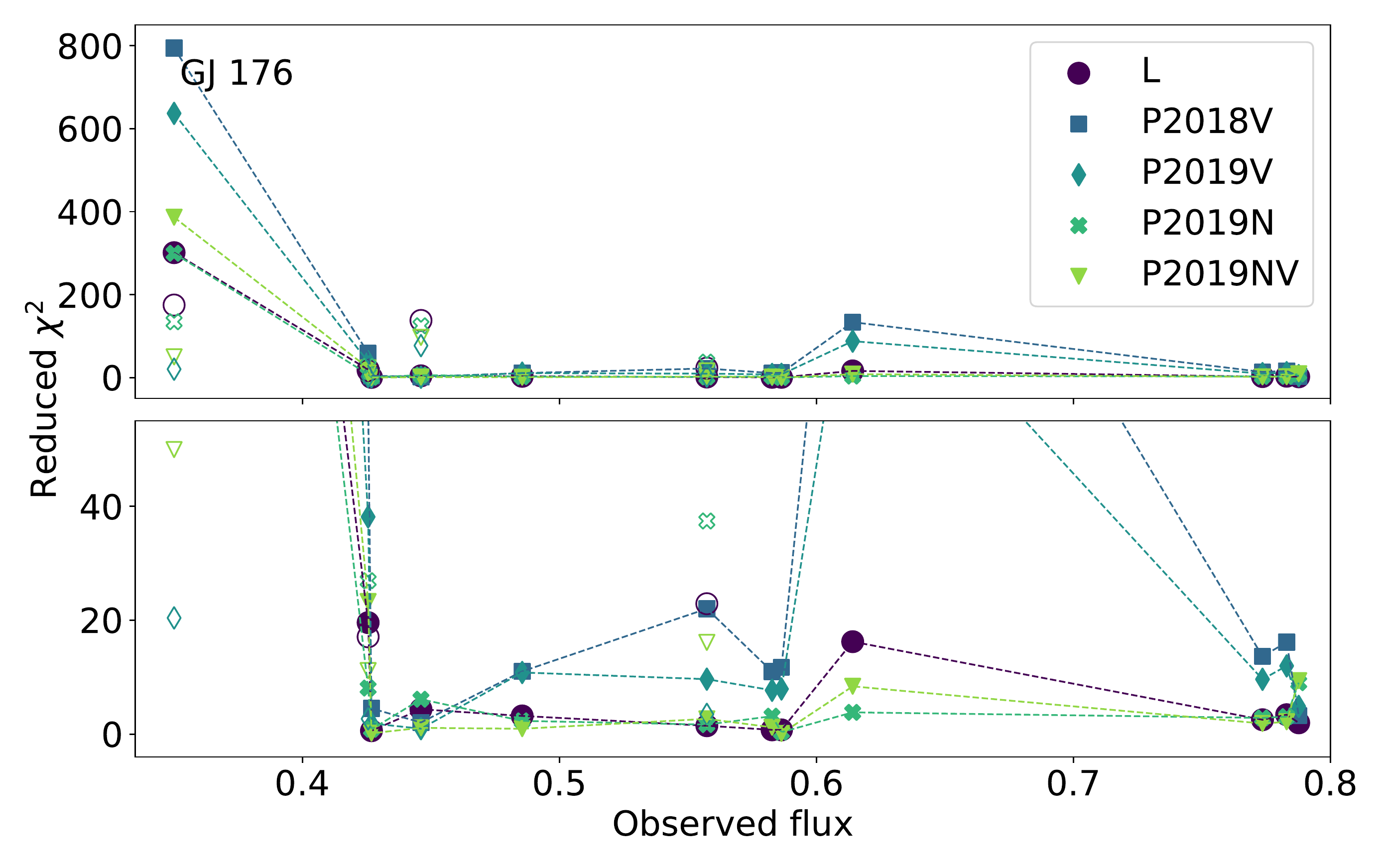} 
   \caption{Reduced $\chi^2$ between the synthetic and observed spectra for the cores of the twelve strongest lines for the star GJ~176 as a function of the observed flux in the centre of the lines (filled symbols). The symbols are the same as in Fig.~\ref{AngTeffDiff} and refer to the parameters used for the synthetic spectra. The upper panel shows the whole range of $\chi^2$ while the lower panel focuses on $\chi^2$ below 50. The open symbols indicate the $\chi^2$ of K lines generated in NLTE for L and P2019. The identifications of all lines are presented in Table~\ref{Tab:Chi2lines} together with the wavelengths and observed fluxes.}
              \label{Fig:GJ176Chi2}%
              \end{center}
\end{figure}

\begin{figure}
   \begin{center}
   \includegraphics[width=9cm]{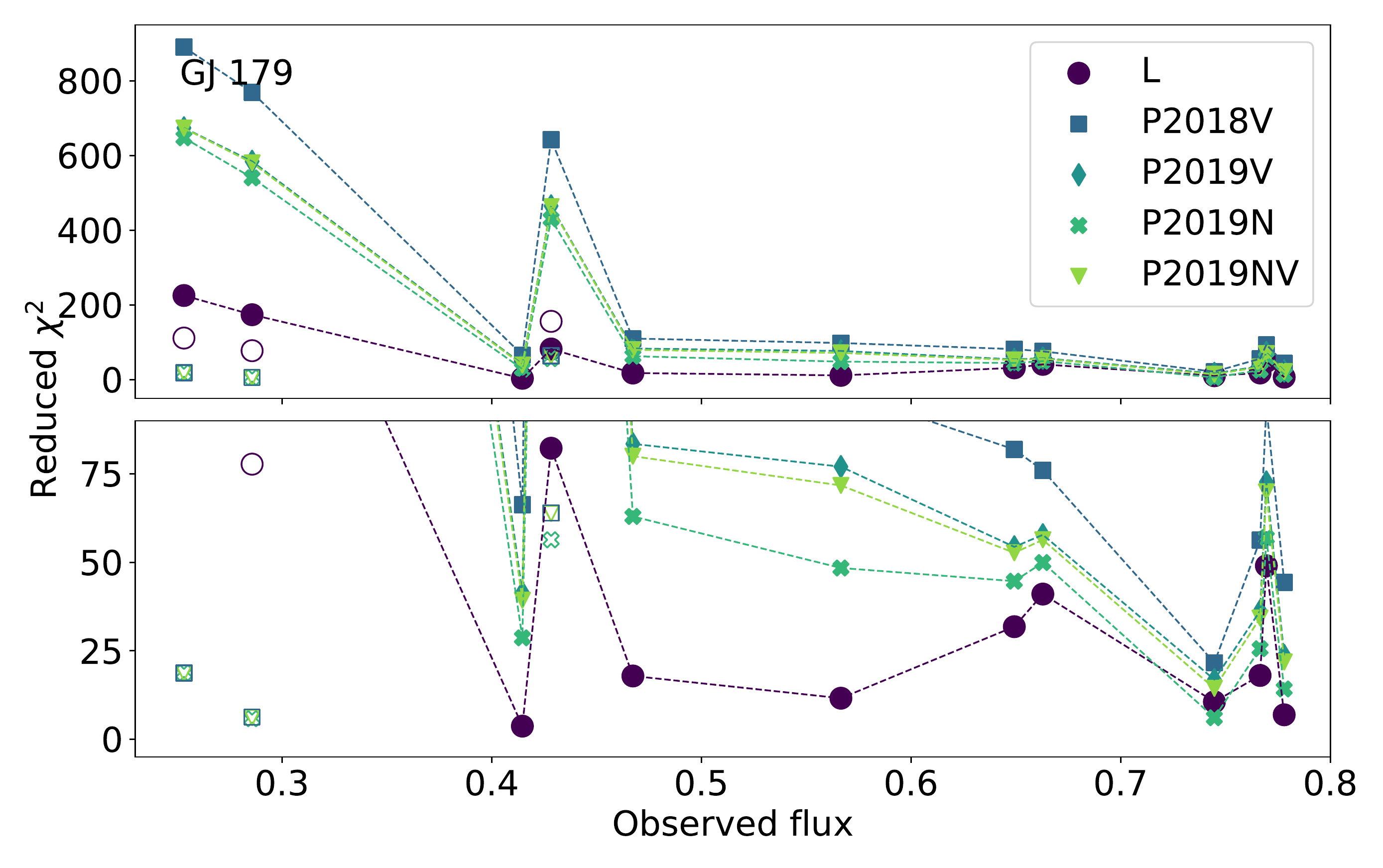} 
   \caption{Same as Fig.~\ref{Fig:GJ176Chi2} for the star GJ~179. The lower panel focuses on $\chi^2$ below 90.}
              \label{Fig:GJ179Chi2}%
              \end{center}
\end{figure}

\begin{figure}
   \begin{center}
   \includegraphics[width=9cm]{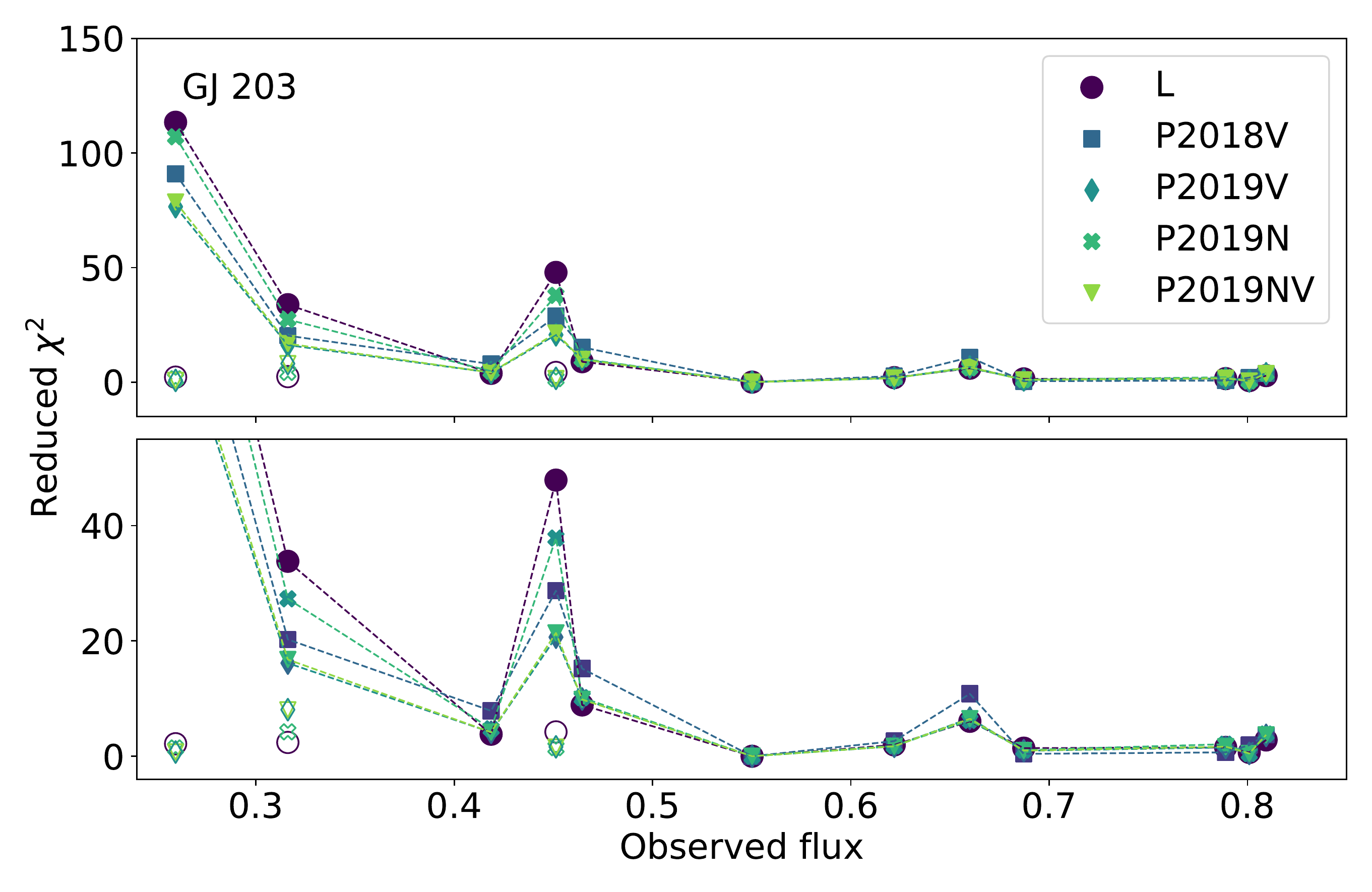} 
   \caption{Same as Fig.~\ref{Fig:GJ176Chi2} for the star GJ~203.}
              \label{Fig:GJ203Chi2}%
              \end{center}
\end{figure}

\begin{figure}
   \begin{center}
   \includegraphics[width=9cm]{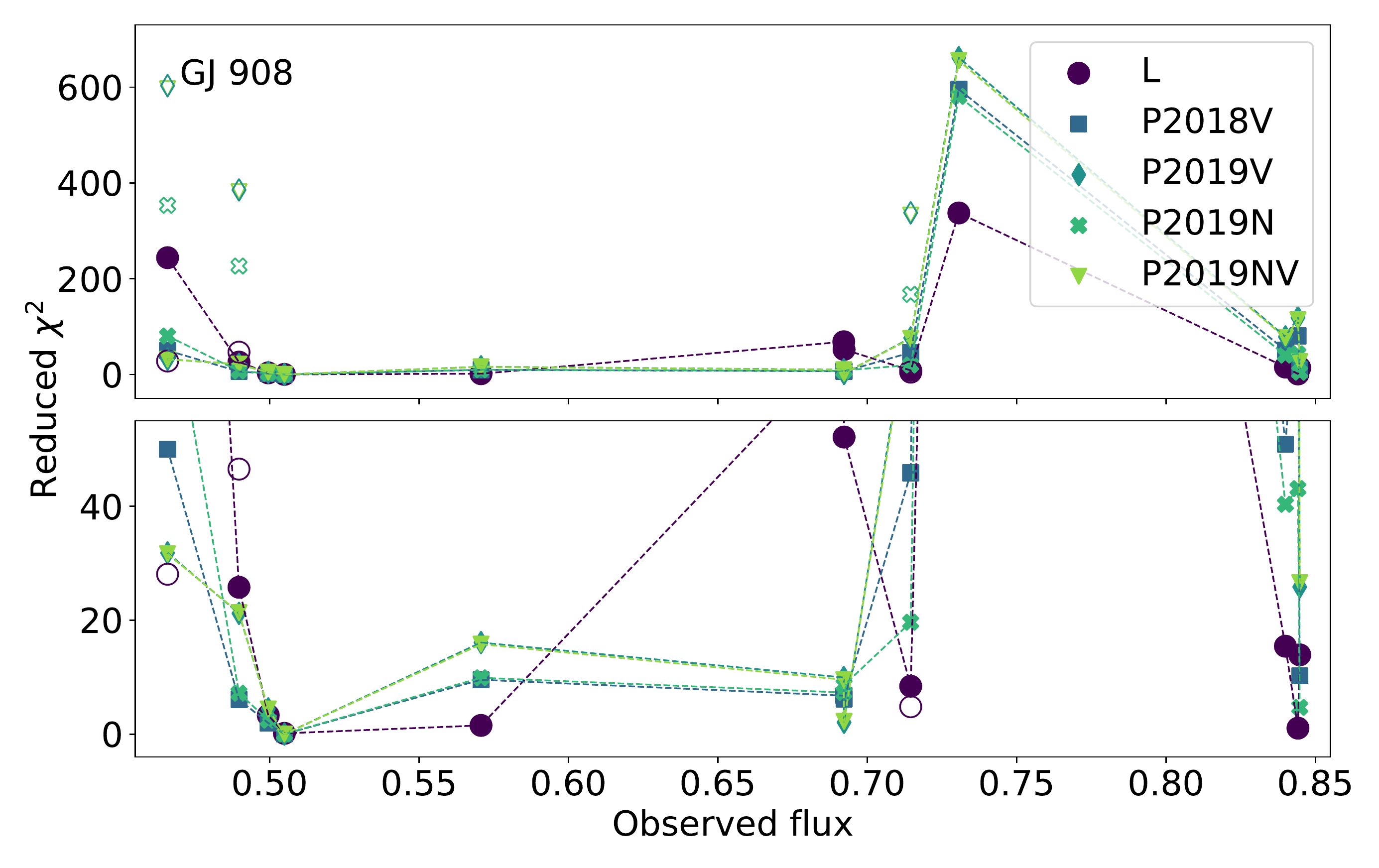} 
   \caption{Same as Fig.~\ref{Fig:GJ176Chi2} for the star GJ~908.}
              \label{Fig:GJ908Chi2}%
              \end{center}
\end{figure}

A $\chi^2$ comparison of the cores of the strongest lines in the synthetic and observed spectra was performed. A small wavelength region of $\pm$0.1~\AA \ around the minimum flux was investigated. The resulting values of reduced $\chi^2$ for four of the stars and the twelve strongest lines are presented in Figs.~\ref{Fig:GJ176Chi2} to \ref{Fig:GJ908Chi2}. The central wavelengths of these lines and the observed fluxes at these wavelengths are listed in Table~\ref{Tab:Chi2lines}. The stars shown in the figures were selected to represent the different cases of agreement of the metallicities from L2016 and L2017 and P2018 and P2019 as follows. Two of the stars (GJ~179 and GJ~908) have metallicity differences larger than the uncertainties, with opposite signs. GJ~203 has agreeing metallicities and a small spread, while GJ~176 has on average agreeing metallicities but with a large spread. In the figures we also show the reduced $\chi^2$ calculated in the same way between the observed and synthetic spectra, but generated with potassium in NLTE for L2016, L2017, and P2019 parameters (open symbols). We did not calculate the reduced $\chi^2$ for P2018 in NLTE since the method was improved in P2019 (see Sect.~\ref{sect:models}). The straight lines in the figures are a visual aid and should not be seen as a prediction of intermediate values. In the upper panels we show the entire range of the reduced $\chi^2$ and in the lower panels we focus on low values of $\chi^2$. The y-scale is the same for all stars in the lower panels, except for GJ~179 which in general has larger $\chi^2$ values. 

The metallicities for GJ~176 from L2016 and P2019 agree within uncertainties and have a large spread. In Fig.~\ref{Fig:GJ176Chi2} we see a spread in $\chi^2$ for several lines. For GJ~179 the derived metallicities are outside of each other's uncertainties and have a large spread. The reduced $\chi^2$ values in Fig.~\ref{Fig:GJ179Chi2} are generally larger than for any other star in this sample, and this star was given as an outlier in previous sections. In Fig.~\ref{Fig:GJ203Chi2} (GJ~203) we see a low and very small spread in the reduced $\chi^2$ for all non-potassium lines. The metallicities derived for this star also have a small spread and agree within uncertainties. GJ~908 has derived metallicities that do not agree within the uncertainties and is a clear outlier in Fig.~\ref{PlotParam}. This star also has a spread in reduced $\chi^2$, which is not as large as for GJ~179. The $\chi^2$ values for P2019 are smaller than for P2018 which indicates an improvement in the method between P2018 and P2019. For GJ~176 and GJ~179 the case of L2016 and L2017 has a lower $\chi^2$ than P2018 and 2019 for most of the lines in \fig{Fig:GJ176Chi2} and \fig{Fig:GJ179Chi2}. For GJ~908 L2017 has a better fit except for two Ti lines in the middle of \fig{Fig:GJ908Chi2}. L2017 and both P2018 and P2019 agree for GJ~203.

At this point we can state that there is an inconsistency between the parameters derived by the different authors, since we cannot reproduce a good fit to the observations with the method of L2016 and L2017 using parameters derived by P2018 and P2019. The parameters are clearly model dependent. However, we cannot say which of them are more realistic because of the various assumptions made in the comparison, not the least the abundances of individual elements used when calculating the synthetic spectra. A change in the abundances could improve or worsen the fit to the observations for any of the parameter sets.

We can see that the majority of the lines with the largest $\chi^2$ are \ion{K}{I} lines. For most of the \ion{K}{I} lines, the $\chi^2$ improves when using NLTE, although there are some exceptions. One is for the star GJ~176 (Fig.~\ref{Fig:GJ176Chi2}) where the fourth and sixth lines are K lines (with observed flux 0.446 and 0.557, respectively) for which LTE has a lower reduced $\chi^2$ than NLTE. When visually comparing these lines we see that the synthetic spectra generated in LTE show a better fit against the observed spectrum than the NLTE spectra. Another exception is the fourth line in Fig.~\ref{Fig:GJ179Chi2} (observed flux 0.428) where the L2017 line calculated in NLTE has a $\chi^2$ about twice as high as the respective LTE line. This line can be seen to the left in Fig. \ref{K_LTEvsNLTE}. All K lines improved when using P2019 parameters with NLTE for this star. The star GJ~908 also shows lines that have a worse fit with K in NLTE. The L2017 line for this star that has a bad fit (observed flux 0.490) is blended with Fe which can cause discrepancies. All K lines generated with parameters from P2019 show a worse fit in NLTE than in LTE for GJ~908.

   \begin{table*}[t]
    \caption{Twelve strongest lines shown in Figs.~\ref{Fig:GJ176Chi2} to \ref{Fig:GJ908Chi2}. The central wavelength and the observed flux at that wavelength are given.}
       \label{Tab:Chi2lines}
       $$
       \begin{tabular}{p{0.05\linewidth}c c c c c c c c c c}
       \hline
       \hline
       \noalign{\smallskip}
        &  &\multicolumn{4}{c}{Observed flux} \\
        Species & $\lambda$ [\AA] & GJ 176 & GJ 179 & GJ 203 & GJ 908\\
       \noalign{\smallskip}
       \hline
       \noalign{\smallskip}
       \ion{Mg}{I} & 11828.2 & 0.486 & ... & 0.550 & 0.505\\
       \ion{K}{I}  & 11690.2 & 0.350 & 0.286 & 0.316 & 0.490\\
       \ion{K}{I}  & 11769.6 & 0.557 & 0.428 & 0.451 & 0.715\\
       \ion{K}{I}  & 11772.8 & 0.446 & 0.253 & 0.260 & 0.466\\
       \ion{K}{I}  & 12522.1 & 0.426 & ... & ... & ...\\
       \ion{Ca}{I} & 13033.6 &  ... & 0.745 & 0.789 & 0.844 \\
       \ion{Ti}{I} & 11780.5 & 0.774 & 0.778 &... & 0.840\\
       \ion{Ti}{I} & 11797.2 & 0.783 & 0.766 & 0.809 & ...\\
       \ion{Ti}{I} & 11892.9 & ... & 0.663 & 0.660 & 0.692\\
       \ion{Ti}{I} & 11949.5 & 0.614 & 0.649 & 0.687 & 0.692\\
       \ion{Ti}{I} & 11973.8 & 0.583 & ... & ... & ...\\
       \ion{Ti}{I} & 13011.9 & ... & 0.769 & 0.801 & 0.845\\
       \ion{Fe}{I} & 11783.3 & 0.583 & 0.567 & 0.622 & 0.731\\
       \ion{Fe}{I} & 11882.8 & ... & 0.415 & 0.419 & 0.500\\
       \ion{Fe}{I} & 11884.0 & ... & 0.467 & 0.465 & 0.571\\
       \ion{Fe}{I} & 11973.0 & 0.427 & ... & ... & ...\\
       \ion{Si}{I} & 12031.5 & 0.788 & ... & ... & ...\\
       
       \hline
       \end{tabular}
       $$
    \tablefoot{The \ion{K}{I} line at 11690.2~\AA \ is blended with a \ion{Fe}{I} line.}
       
   \end{table*}

\subsection{Comparison of model atmospheres}
\label{sect:models}

   \begin{figure}
   \centering
   \resizebox{\hsize}{!}{\includegraphics{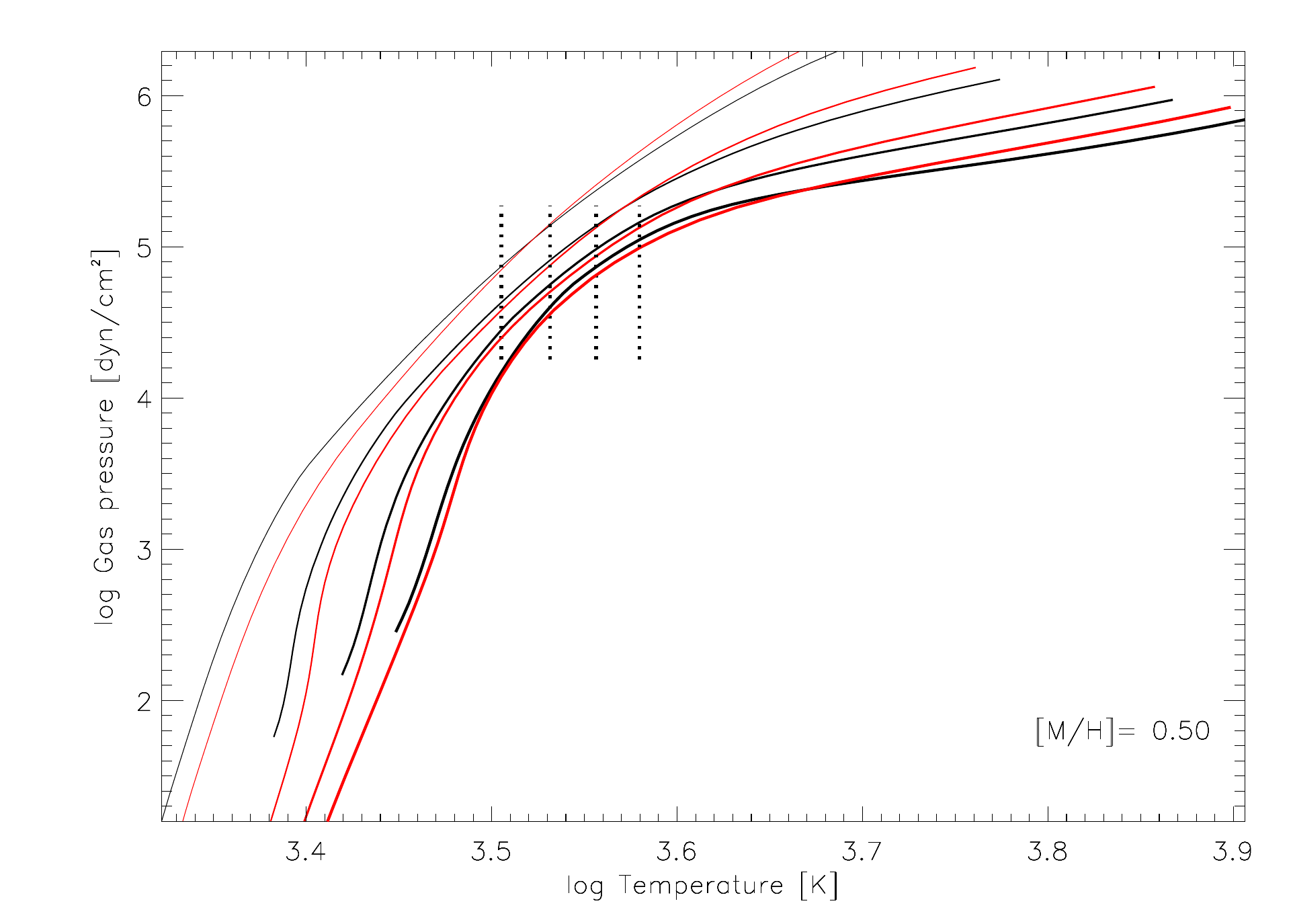}}
   \resizebox{\hsize}{!}{\includegraphics{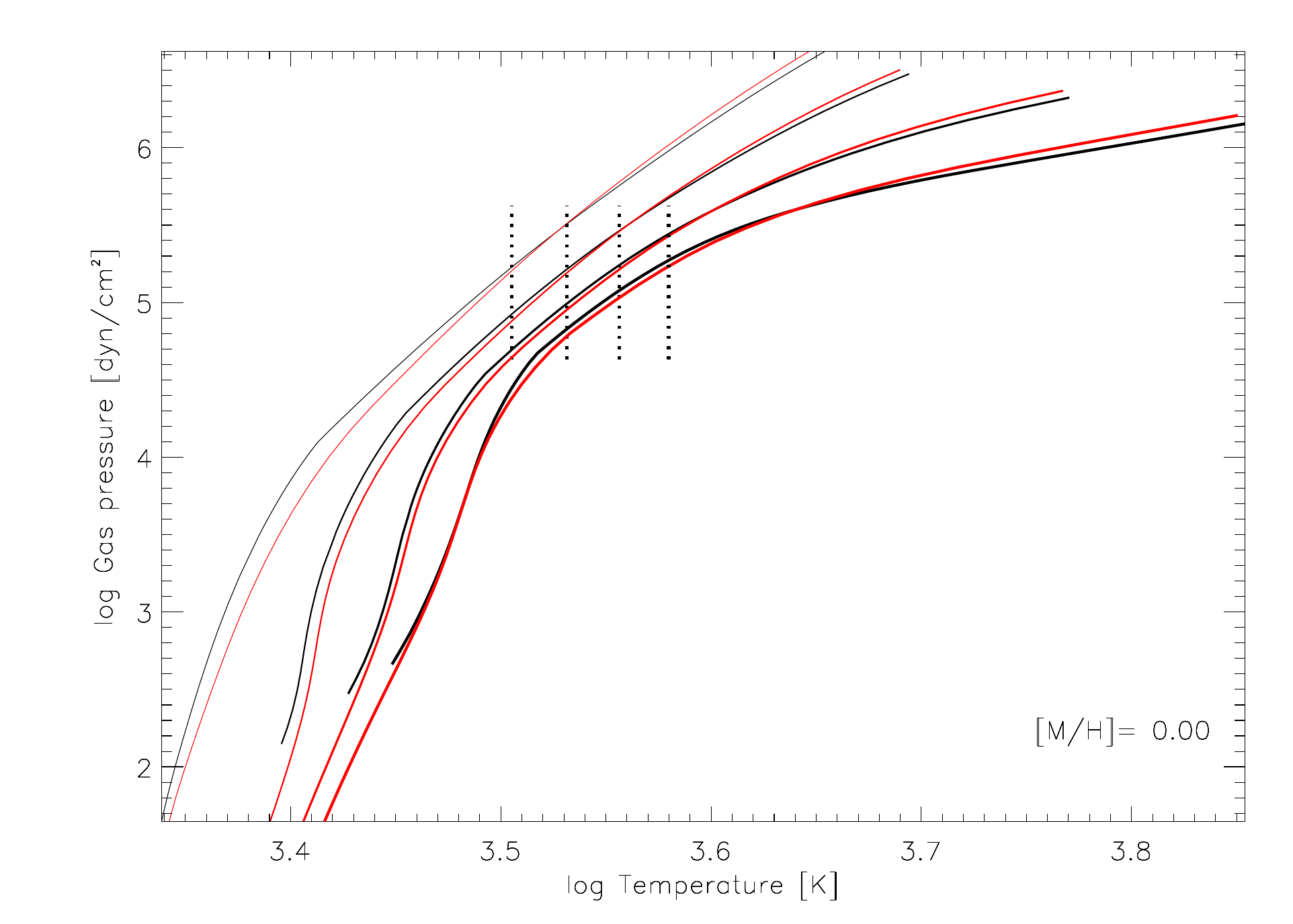}}
   \resizebox{\hsize}{!}{\includegraphics{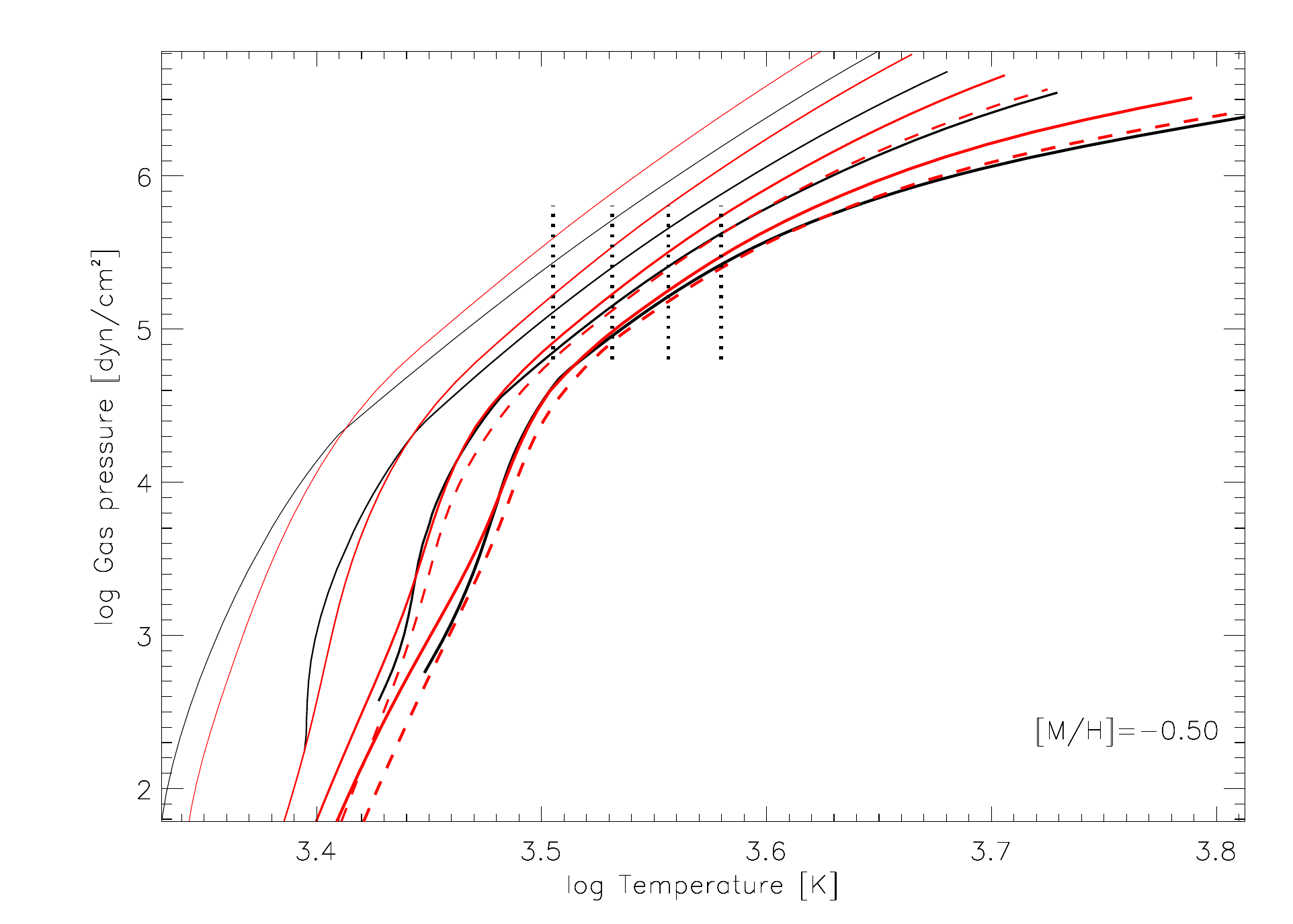}}
   \caption{Comparison of temperature-pressure profiles for \marcs{} (black solid lines) and \phoenix{} (red solid lines) model atmospheres used in L2016 and L2017 and P2018 and P2019  studies, respectively (see text for details). Each panel shows profiles for $T_{\rm eff}$=3200~K, 3400~K, 3600~K, and 3800~K (from left to right). Vertical dotted lines corresponding to the values of the effective temperatures are also shown. The three panels show profiles for three different metallicities (decreasing from the top to the bottom panel). The red dashed lines in the bottom panel correspond to \phoenix{} models with [$\alpha$/Fe] = +0.2 (for $T_{\rm eff}$=3600~K and 3800~K).}
              \label{fig:models}
    \end{figure}

In order to separate the discrepancies in derived parameters caused by NLTE effects and by using different atmospheric models we compare the model structures used by L2016, L2017, P2018, and P2019. L2016 and L2017 used \marcs{} atmospheric models \citep{Gustafsson2008A&A}, while P2018 and P2019 based their analysis on \phoenix{} models. 
The \phoenix{} models used by P2018 and P2019 were of two different flavours, mainly differing in the equation of state (EOS) and the atomic and molecular line lists used.
The models used by P2018 are presented in \citet{Husser2013PHOENIX-ACES} and are made available on-line (see below). They use the ACES\footnote{Astrophysical Chemical Equilibrium Solver} EOS with thermodynamic data for hundreds of species, including about 250 molecules and over 200 condensates.
The models applied by P2019 use the SESAM\footnote{Stoichiometric Equilibrium Solver for Atoms and Molecules} EOS, which is discussed extensively in \citet{Meyer17}. These models are not (yet) publicly available.
Both ACES and SESAM calculate the chemical equilibrium based on the Villars-Cruise-Smith method \citep{smith1982chemical}. 
The SESAM EOS was developed in order to enable the \phoenix{} code to model Earth-like planetary atmospheres with effective temperatures of a few hundred K. This included an improved treatment of condensed species, together with improvements on the numerical side.

\citet{Meyer17} compared \phoenix{} models calculated with both the ACES and the SESAM EOS for an effective temperature of 3000~K and concluded that both versions resulted in very similar atmospheric structures. A comparison of spectra calculated for M-dwarf parameters with \phoenix{} using ACES and SESAM is provided in Fig.~2 of P2019. Deviations are mostly seen for individual atomic (\ion{Mg}{i}, \ion{Ca}{ii}) and molecular (e.g. TiO) lines, which indicates that the differences are mainly due to different atomic and molecular data used. The model structures for M~dwarfs are expected to be very similar for the two \phoenix{} versions, given that the implemented improvements focused on a much lower temperature range.

Based on the above, we use the publicly available \phoenix{} version (with the ACES EOS) for a comparison with \marcs{} models. In the computation of the \marcs{} models, a Newton-Raphson method was used to calculate the chemical equilibrium, including about 500 molecular species \citep{Gustafsson2008A&A}. We note that when synthetic spectra based on \marcs{} models are calculated by \sme{}, \sme{} uses its own EOS to re-calculate the chemical equilibrium and the partial pressures, including electron pressure (also based on a Newton-Raphson method, see \citealt{sme2_2017}), for atoms, ions, and about 200 molecular species. 

Both of the atmospheric model codes use the assumption of local thermodynamic equilibrium (LTE). In the case of \phoenix{} there is an option to use line profiles calculated in non-LTE for some species (\ion{Li}{i}, \ion{Na}{i}, \ion{K}{i}, \ion{Ca}{i}, \ion{Ca}{ii}). However, this option was only used for models with $T_{\rm eff} \ge$ 4000~K, according to \citet{Husser2013PHOENIX-ACES}.

One of the differences between \marcs{} and \phoenix{} models is given by the abundances of elements assumed in the model calculations. In the case of \marcs{} the reference solar abundance mixture is that of \citet{2007SSRv..130..105G}. 
Furthermore, the variation of abundances as a function of metallicity used in the model grid examined here reflects the typical elemental abundance ratios in stars in the solar neighbourhood.
Specifically, [$\alpha$/Fe] increases from 0.0 at solar metallicity to +0.1 at [M/H]=$-0.25$ and +0.2 at [M/H]=$-0.5$, and further to +0.3 at [M/H]=$-0.75$ and +0.4 at [M/H]=$-1.0$.
In the \phoenix{} models the reference solar abundances are taken from \citet{2009ARA&A..47..481A} in P2018, 
complemented by results from \citet{2011SoPh..268..255C} in P2019. 
\phoenix{} models with a range of [$\alpha$/Fe] values are available for metallicities equal to or less than 0.0, but only for $T_{\rm eff} \ge$ 3500~K. P2018 and P2019 exclusively used models with [$\alpha$/Fe]=0 in their analysis (see Sect.~3 in P2018).

The \marcs{} models used for the comparison are those included in the \sme{} package, which were used in L2016 and L2017.
The model data provided are the mass column density in units of g\,cm$^{-2}$, the temperature in units of K, the electron number density and the atomic number density in units of cm$^{-3}$, the density in units of g\,cm$^{-3}$, and the optical depth at a reference wavelength of 5000~\AA.

The \phoenix{} models used for the comparison were downloaded from the \emph{Göttingen Spectral Library}\footnote{\url{http://phoenix.astro.physik.uni-goettingen.de/?page_id=108}, accessed June 2020} described in \citet{Husser2013PHOENIX-ACES}. The on-line documentation states that for each atmospheric model the optical depth, the temperature in units of K, the gas pressure in units of dyn\,cm$^{-2}$, the density in units of g\,cm$^{-3}$, and the electron partial pressure in units of dyn\,cm$^{-2}$ is provided. However, the downloaded data files do not contain the electron pressure. The optical depth in the \phoenix{} data files is given at a reference wavelength of 12\,000~\AA, according to \citet{Husser2013PHOENIX-ACES}.
Thus, there are three quantities in common that can be compared: temperature, gas pressure, and density.

Figure~\ref{fig:models} shows the gas pressure profiles of \marcs{} and \phoenix{} model atmospheres as a function of temperature, for four different effective temperatures and three different metallicities. The surface gravity is 4.5~dex in all cases.
In the lower part of the atmosphere the gas pressure is higher in \phoenix{} models than in \marcs{} models at the same temperature, and the differences increase towards larger depths. The opposite is the case in the upper part of the atmosphere.
The difference in density profiles shows a very similar behaviour.
The differences increase towards smaller $T_{\rm eff}$ values, with maximum differences of about 0.3~dex (corresponding to a factor 2).
Using $\log g$=5.0 instead of $\log g$=4.5 results in slightly smaller differences. 
The differences in gas pressure at the location where the temperature is equal to the effective temperature, corresponding to the line-forming region, are small. They range from being insignificant at $T_{\rm eff}$=3200~K to about 15\% at $T_{\rm eff}$=3800~K (0.06~dex).

The differences are smallest at solar metallicity. They are slightly larger at [M/H]=$+0.5$ and even larger at [M/H]=$-0.5$, in particular in the lower part of the atmosphere, including the line-forming region. However, most of the difference at [M/H]=$-0.5$ seems to be due to the different abundances of $\alpha$-elements used in the models. As mentioned above \phoenix{} models with [$\alpha$/Fe] = +0.2 are available for $T_{\rm eff} \ge$ 3500~K.
As can be seen in Fig.~\ref{fig:models} (bottom panel) the profiles for these models follow closely those of the corresponding \marcs{} models, except in some parts of the upper atmosphere, above the line-forming region.

\section{Conclusions} \label{conclusion}

In recent years many studies deriving M-dwarf atmospheric parameters have been published.  We compared parameters derived from high-resolution optical and near-infrared spectra by L2016, L2017, P2018, P2019, and R2018 with each other for an overlapping sample of stars. In the case of effective temperature we also compared with effective temperatures derived from interferometric diameters and bolometric fluxes. We find that the effective temperatures generally agree, although the temperatures from P2018 and P2019 are often higher than those from L2016, L2017, and \citet{Rabus2019}. R2018 agrees with L2016, L2017, P2018, and P2019 but shows a larger spread compared to \citet{Rabus2019}. For the surface gravity we see a larger spread. The P2018 surface gravities are higher than those from L2016 and L2017 while the ones from P2019 are lower. R2018 surface gravities are higher than all other studies in most cases. The metallicity also shows a spread, where P2018 and P2019 metallicities are grouped around solar metallcities. Furthermore, the metallicities from P2019 obtained in the near-infrared are higher than the others. Metallicities from R2018 are spread over the whole parameter space and show no correlation with L2016, L2017, P2018, nor P2019. The large discrepancy between R2018 parameters and what was derived in other studies can be explained by degeneracies between the metallicity and surface gravity. We identified some outliers, for example the star GJ~908 for which the effective temperature and surface gravity from L2017 and P2019 agree within uncertainties while the metallicities differ significantly. R2018 derived a metallicity 1 dex higher than was derived by L2017 for this star.

We investigated the contribution of NLTE effects to the difference in derived parameters. We generated synthetic spectra in LTE and NLTE using grids of departure coefficients for \ion{Fe}{I} and \ion{K}{I} for the NLTE spectra and compared these to each other. For iron we found the difference between LTE and NLTE to be insignificant. For potassium the difference was larger. The largest difference was observed for lines with the lowest excitation energies. We quantified the effect by determining abundance corrections for \ion{K}{I} lines in the near-infrared that were most affected (with the addition of the resonance line at 7699~\AA). The largest abundance corrections of around $-0.2$~dex were found for the low-excitation lines in the warmest stars. Spectroscopically derived stellar parameters such as effective temperature and metallicity are also affected by NLTE effects. We found that LTE underestimates $T_{\rm eff}$ while it overestimates the metallicity. The largest difference in $T_{\rm eff}$ was found for one of the warmest stars in the sample, GJ~880, which shows a difference of $213$~K. The coolest star with the highest overall metallicity (GJ~179) shows the largest difference in metallicity, $-0.24$~dex.

This difference in metallicity between LTE and NLTE could possibly partly explain why the metallicities derived in the near-infrared by P2019 are higher than the metallicities derived in the visual. The difficulties with fitting \ion{K}{I} lines in the near-infrared reported by P2019 could also be explained by NLTE effects\footnote{P2018 and P2019 removed K lines which showed a bad fit (V.~Passegger, private communication). This would mitigate some of the biases in stellar parameters arising from not accounting for NLTE effects in potassium.}.

We generated synthetic spectra with the parameters from L2016, L2017, P2018, and P2019 for four of the stars in the sample and assessed how well these reproduced observed near-infrared spectra through a $\chi^2$ analysis. R2018 was excluded due to the possible degeneracy between the parameters. 
For all of the parameters we found large differences for the strongest lines of potassium. Strong lines for other elements in general show a better fit. However, two of the stars that were identified as outliers when comparing the metallicity between L2016 and L2017 and P2018 and P2019 (GJ~179 and GJ~908) show high $\chi^2$-values for non-potassium lines. This indicates a problem with either the models and/or the derived parameters. Synthetic spectra generated using NLTE for \ion{K}{I} improved the $\chi^2$-values for most of the potassium lines, except for the star GJ~176. An explanation for the K lines that are worse with P2019 parameters in NLTE in all stars (except GJ~203) could be that the parameters were derived using LTE. This means that the best-fit parameters to some degree compensate for the NLTE effects. When these parameters then are used in NLTE they show a worse fit. This, however, can not explain the difference seen for the lines generated with parameters from L2016 and L2017, since K lines were excluded in that study. Another possible explanation is related to non-solar chemical abundances. In particular, for GJ~176 the potassium abundance in the atmosphere could be smaller than the solar abundance that was assumed in our analysis. This would weaken the K lines and hence give a better fit for lines generated in LTE. As can be seen in Fig.~\ref{K_LTEvsNLTE} the lines are weaker in the LTE spectra than in the NLTE spectra.

The atmospheric models used by P2018, P2019, L2016, and L2017 agree fairly well, with most of the differences occurring near the surface or deep in the interior. The largest contribution to differences in the results of the spectroscopic analyses is probably found for metal-poor stars, due to differences in the assumed [$\alpha$/Fe] abundances.
We also found that the difference in atmospheric models in the line-forming region increases towards higher effective temperatures. Since the NLTE effects of K also increase towards higher temperatures it is possible that these effects amplify each other. 
These effects could explain why the star GJ~908, which was identified as an outlier in Sect.~\ref{sect:StellarParam}, shows a large difference in metallicity. This star has a low metallicity and a rather high effective temperature. On the other hand, the star GJ~203 has a high surface gravity, corresponding to small model differences, and a fairly low effective temperature, and for this star the parameters of L2017 and P2018 and P2019 agree.

Another star that was given as an outlier in Sect.~\ref{sect:StellarParam} was GJ~179 which has a low effective temperature indicating a small model difference and a high metallicity which indicates a slightly larger model difference. If we look at the leftmost curves in the top panel of Fig.~\ref{fig:models} we see that the differences between \marcs{} and \phoenix{} models are very small. Neither NLTE effects nor model difference can fully explain the difference in derived metallicities for this star. We also see that this star has high $\chi^2$-values for many lines, not only potassium lines (see Fig.~\ref{Fig:GJ179Chi2}). A possible explanation could again be non-solar abundances. High-resolution spectroscopic observations of this star covering a large wavelength range are needed to do derive abundances of individual elements. GJ~849 was also identified as an outlier in section \ref{sect:StellarParam}. This star has similar effective temperature and metallicity as GJ~179 when looking at the parameters derived by L2016 and L2017 and could therefore be affected by NLTE in the same way. However, L2016 derived a significantly lower effective temperature than P2018 and P2019, regardless of wavelength range.

Another major source of uncertainty for the derived parameters is the quality of the atomic data in the line list.
We did not compare the atomic data used by L2016, L2017, P2018, P2019, and R2018 since the line list from neither P2018 nor P2019 was available. 
However, a careful assessment of the available atomic data in the near-infrared should be done and the most precise and accurate data should be selected for any future analysis of high-resolution spectra. In addition, the magnetic sensitivity of the lines should be taken into account in order to assess the influence of magnetic fields on the derived parameters.

In conclusion, the cores of the strongest \ion{K}{I} lines in the near-infrared are clearly affected by NLTE and this needs to be taken into account in any analysis of high-resolution M-dwarf spectra in order to derive the most realistic parameters.
We recommend that NLTE is used for the calculation of model spectra or that the lines are avoided. \ion{Fe}{I} can be calculated in LTE since the difference between LTE and NLTE are insignificant. NLTE effects of other elements remain to be investigated for M~dwarfs. For example, some \ion{Ca}{I} lines show a discrepancy between synthetic and observed spectra for some of the stars investigated here (see Fig.~\ref{Tefffree}). Apart from inadequate atomic data, NLTE effects could provide a possible explanation since \ion{Ca}{I} shows NLTE effects in FGK stars \citep{2017A&A...605A..53M,2020A&A...637A..80O}.

In order to validate stellar parameters derived by spectroscopic methods we need model-independent stellar parameters to compare with, such as $\teff$ from interferometry. Further research into asteroseismology for M~dwarfs is also needed so that $\log g$ can be constrained without using empirical calibrations. With a well-defined set of fully characterised M-dwarf benchmark stars we will be able to disentangle the contribution of NLTE effects to uncertainties in the spectroscopic analysis from other aspects, such as the atmospheric model structure, atomic data, and non-solar abundances of individual elements.  

\begin{acknowledgements}
      We thank Anish Amarsi for providing the NLTE grid for potassium and for fruitful discussions. We acknowledge support from the Swedish National Space Agency (SNSA/Rymdstyrelsen) and the Swedish Research Council (Vetenskapsr{\AA}det). This work has made use of the VALD database, operated at Uppsala University, the Institute of Astronomy RAS in Moscow, and the University of Vienna.
\end{acknowledgements}

\bibliographystyle{aa}
\bibliography{MdwarfComp,TereseOlander.046302,T3000g4_10000-16000}

\begin{appendix}
\section{Derived parameters}

This appendix provides a table with the stellar atmospheric parameters derived by L2016, L2017, P2018, P2019, and R2018, referred to in Sect.~\ref{sect:StellarParam}.

\longtab[1]{
\begin{longtable}{l l l c c c c c}
\caption{Atmospheric stellar parameters from \citet{Lind2016}, \citet{Lind2017}, \citet{Pass2018,Pass2019}, and \citet{Rajpurohit2018CARMENES} (references L2016, L2017, P2018, P2019, and R2018 in column ``Ref.'', respectively) for the stars in the sample. The letters V, N, and NV in column ``Ref.'' indicate the wavelength regions (visual, near-infrared, and both combined) which P2018, P2019, and R2018 used to derive the parameters. R2018 has no specified $\varv\sin i$, $\varv_{mic}$, nor $\varv_{mac}$.}\label{Param}\\
\hline
\hline
\noalign{\smallskip}
Star & Ref. & $\rm T_{eff}$  & $log\ g$  & [M/H]  & $\varv\sin i$ & $\varv_{mic}$ & $\varv_{mac}$\\
 & & [K] & $[cm s^{-2}]$ & &  $[km s^{-1}]$ & $[km s^{-1}]$ & $[km s^{-1}]$\\
\hline
\endfirsthead
\caption{Continued.} \\
\hline
\noalign{\smallskip}
Star & Ref. & $\rm T_{eff}$  & $log\ g$  & [M/H]  & $\varv\sin i$ & $\varv_{mic}$ & $\varv_{mac}$\\
 & & [K] & $[cm s^{-2}]$ & &  $[km s^{-1}]$ & $[km s^{-1}]$ & $[km s^{-1}]$\\
\hline
\endhead
\hline
\endfoot
\hline
\endlastfoot
\noalign{\smallskip}
       \multirow{5}{*}{1 GJ176}& P2018V  & 3582$\pm$ 51  & 4.88$\pm$ 0.07 & -0.08$\pm$ 0.16 & 3.00 & 0.25 & 0.50\\
                               & P2019V  & 3520$\pm$ 51  & 4.79$\pm$ 0.04 & -0.08$\pm$ 0.16 & 2.00 & 0.25 & 0.50\\
                               & P2019N  & 3654$\pm$ 56  & 4.66$\pm$ 0.04 & +0.48$\pm$ 0.16 & 2.00 & 0.25 & 0.50\\
                             & P2019NV & 3689$\pm$ 54  & 4.66$\pm$ 0.06 & +0.33$\pm$ 0.19 & 2.00 & 0.25 & 0.50\\
                             & R2018NV & 3500$\pm$ 100 & 5.10$\pm$ 0.30 & 0.00$\pm$ 0.30 & ... & ... & ... \\ 
                             & L2016  & 3550$\pm$ 100 & 4.76$\pm$ 0.08 & +0.11$\pm$ 0.09 & 1.00 & 1.00 & 2.01\\
\hline
\noalign{\smallskip}                            
        \multirow{5}{*}{2 GJ179}& P2018V  & 3391$\pm$ 51 & 5.00$\pm$ 0.07 &  0.00$\pm$ 0.16 & 2.50 & 0.20 & 0.40 \\
                             & P2019V  & 3339$\pm$ 51 & 4.85$\pm$ 0.04 & +0.02$\pm$ 0.16 & 2.00 & 0.20 & 0.40 \\
                             & P2019N  & 3349$\pm$ 56 & 4.80$\pm$ 0.04 & +0.14$\pm$ 0.16 & 2.00 & 0.20 & 0.40 \\
                             & P2019NV & 3341$\pm$ 54 & 4.84$\pm$ 0.06 & +0.04$\pm$ 0.19 & 2.00  & 0.20 & 0.40\\
                             & R2018NV & 3300$\pm$ 100 & 5.00$\pm$ 0.30 & +0.30$\pm$ 0.30 & ... & ... & ... \\ 
                             & L2017  & 3300$\pm$ 100 & 4.89$\pm$ 0.10 & +0.36$\pm$ 0.04 & 1.00 & 0.25 & 0.06 \\
        \hline
       \noalign{\smallskip}                    
       \multirow{5}{*}{3 GJ203}& P2018V  & 3362$\pm$ 51 & 5.03$\pm$ 0.07 & -0.03$\pm$ 0.16 & 3.00 & 0.25 & 0.50\\
                             & P2019V  & 3332$\pm$ 51 & 4.86$\pm$ 0.04 & -0.01$\pm$ 0.16 & 2.00 & 0.25 & 0.50\\
                             & P2019N  & 3379$\pm$ 56 & 4.83$\pm$ 0.04 & 0.00$\pm$ 0.16  & 2.00 & 0.25 & 0.50\\
                             & P2019NV & 3335$\pm$ 54 & 4.85$\pm$ 0.06 & 0.00$\pm$ 0.19  & 2.00 & 0.25 & 0.50\\
                             & R2018NV & 3400$\pm$ 100 & 5.10$\pm$ 0.30 & +0.40$\pm$ 0.30 & ... & ... & ... \\
                             & L2017  & 3425$\pm$ 100 & 5.01$\pm$ 0.10 & -0.13$\pm$ 0.04 & 1.00 & 0.25 & 0.02 \\
       \hline
       \noalign{\smallskip}
       \multirow{5}{*}{4 GJ436}& P2018V  & 3512$\pm$ 51 & 4.90$\pm$ 0.07 & -0.02$\pm$ 0.16 & 3.00 & 0.25 & 0.50\\
                             & P2019V  & 3459$\pm$ 51 & 4.79$\pm$ 0.04 & -0.01$\pm$ 0.16 & 2.00 & 0.25 & 0.50\\
                             & P2019N  & 3571$\pm$ 56 & 4.69$\pm$ 0.04 & +0.30$\pm$ 0.16 & 2.00 & 0.25 & 0.50\\
                             & P2019NV & 3472$\pm$ 54 & 4.77$\pm$ 0.06 & +0.03$\pm$ 0.19 & 2.00 & 0.25 & 0.50\\
                             & R2018NV & 3500$\pm$ 100 & 4.00$\pm$ 0.30 & -0.50$\pm$ 0.30 & ... & ... & ... \\
                             & L2016  & 3400$\pm$ 100 & 4.80$\pm$ 0.08 & +0.03$\pm$ 0.06 & 1.00 & 1.00 & 0.08 \\
       \hline
       \noalign{\smallskip}  
       \multirow{5}{*}{5 GJ514}& P2018V  & 3704$\pm$ 51 & 4.82$\pm$ 0.07 & -0.15$\pm$ 0.16 & 3.00 & 0.30 & 0.60 \\
                             & P2019V  & 3720$\pm$ 51 & 4.69$\pm$ 0.04 & +0.07$\pm$ 0.16 & 2.00 & 0.30 & 0.60\\
                             & P2019N  & 3722$\pm$ 56 & 4.67$\pm$ 0.04 & +0.18$\pm$ 0.16 & 2.00 & 0.30 & 0.60\\
                             & P2019NV & 3745$\pm$ 54 & 4.67$\pm$ 0.06 & +0.14$\pm$ 0.19 & 2.00 & 0.30 & 0.60\\
                             & R2018NV & 3700$\pm$ 100 & 5.20$\pm$ 0.30 & -0.40$\pm$ 0.30 & ... & ... & ... \\
                             & L2017  & 3727$\pm$ 100 & 4.78$\pm$ 0.10 & +0.07$\pm$ 0.07 & 1.30 & 0.35 & 0.27 \\
       \hline
       \noalign{\smallskip} 
       \multirow{5}{*}{6 GJ581}& P2018V  & 3430$\pm$ 51 & 5.00$\pm$ 0.07 & -0.09$\pm$ 0.16 & 3.00 & 0.25 & 0.50\\
                             & P2019V  & 3415$\pm$ 51 & 4.85$\pm$ 0.04 & -0.02$\pm$ 0.16 & 2.00 & 0.25 & 0.50\\
                             & P2019N  & 3424$\pm$ 56 & 4.83$\pm$ 0.04 & +0.03$\pm$ 0.16 & 2.00 & 0.25 & 0.50\\
                             & P2019NV & 3413$\pm$ 54 & 4.85$\pm$ 0.06 & -0.02$\pm$ 0.19 & 2.00 & 0.25 & 0.50\\
                             & R2018NV & 3400$\pm$ 100 & 5.00$\pm$ 0.30 & 0.00$\pm$ 0.30 & ... & ... & ... \\
                             & L2016  & 3350$\pm$ 100 & 4.92$\pm$ 0.08 & -0.02$\pm$ 0.13 & 1.00 & 1.00 & 0.33 \\
       \hline
       \noalign{\smallskip}
       \multirow{5}{*}{7 GJ628}& P2018V  & 3378$\pm$ 51 & 5.01$\pm$ 0.07 & +0.01$\pm$ 0.16 & 3.00 & 0.20 & 0.40\\
                             & P2019V  & 3320$\pm$ 51 & 4.75$\pm$ 0.04 & -0.01$\pm$ 0.16 & 2.00 & 0.20 & 0.40\\
                             & P2019N  & 3353$\pm$ 56 & 4.73$\pm$ 0.04 & +0.07$\pm$ 0.16 & 2.00 & 0.20 & 0.40\\
                             & P2019NV & 3305$\pm$ 54 & 4.75$\pm$ 0.06 & +0.01$\pm$ 0.19 & 2.00 & 0.20 & 0.40\\
                             & R2018NV & 3400$\pm$ 100 & 5.00$\pm$ 0.30 & +0.40$\pm$ 0.30 & ... & ... & ... \\
                             & L2016  & 3275$\pm$ 100 & 4.93$\pm$ 0.08 & +0.12$\pm$ 0.14 & 1.00 & 1.00 & 0.04 \\
       \hline
       \noalign{\smallskip}
       \multirow{5}{*}{8 GJ849}& P2018V  & 3454$\pm$ 51 & 4.96$\pm$ 0.07 & -0.01$\pm$ 0.16 & 3.00 & 0.25 & 0.50 \\
                             & P2019V  & 3414$\pm$ 51 & 4.82$\pm$ 0.04 & +0.05$\pm$ 0.16 & 2.00 & 0.25 & 0.50\\
                             & P2019N  & 3633$\pm$ 56 & 4.68$\pm$ 0.04 & +0.54$\pm$ 0.16 & 2.00 & 0.25 & 0.50\\
                             & P2019NV & 3427$\pm$ 54 & 4.80$\pm$ 0.06 & +0.09$\pm$ 0.19 & 2.00 & 0.25 & 0.50\\
                             & R2018NV & 3400$\pm$ 100 & 5.10$\pm$ 0.30 & +0.30$\pm$ 0.30 & ... & ... & ... \\
                             & L2016  & 3350$\pm$ 100 & 4.76$\pm$ 0.08 & +0.28$\pm$ 0.07 & 1.00 & 1.00 & 0.05 \\
       \hline
       \noalign{\smallskip}
       \multirow{5}{*}{9 GJ876}& P2018V  & 3359$\pm$ 51 & 5.01$\pm$ 0.07 & +0.06$\pm$ 0.16 & 2.50 & 0.20 & 0.40\\
                             & P2019V  & 3317$\pm$ 51 & 4.75$\pm$ 0.04 & 0.00$\pm$ 0.16 & 2.00 & 0.20 & 0.40\\
                             & P2019N  & 3286$\pm$ 56 & 4.74$\pm$ 0.04 & +0.05$\pm$ 0.16 & 2.00 & 0.20 & 0.40\\
                             & P2019NV & 3305$\pm$ 54 & 4.75$\pm$ 0.06 & 0.00$\pm$ 0.19 & 2.00 & 0.20 & 0.40\\
                             & R2018NV & 3200$\pm$ 100 & 5.00$\pm$ 0.30 & +0.40$\pm$ 0.30 & ... & ... & ... \\
                             & L2016  & 3250$\pm$ 100 & 4.89$\pm$ 0.08 & +0.19$\pm$ 0.15 & 1.00 & 1.00 & 0.05 \\
       \hline
       \noalign{\smallskip}
       \multirow{5}{*}{10 GJ880}& P2018V  & 3787$\pm$ 51 & 4.70$\pm$ 0.07 & +0.10$\pm$ 0.16 & 2.50 & 0.30 & 0.60\\*
                             & P2019V  & 3789$\pm$ 51 & 4.65$\pm$ 0.04 & +0.32$\pm$ 0.16 & 2.00 & 0.30 & 0.60\\*
                             & 2P019N  & 3784$\pm$ 56 & 4.65$\pm$ 0.04 & +0.53$\pm$ 0.16 & 2.00 & 0.30 & 0.60\\*
                             & P2019NV & 3810$\pm$ 54 & 4.65$\pm$ 0.06 & +0.38$\pm$ 0.19 & 2.00 & 0.30 & 0.60\\*
                             & R2018NV & 3700$\pm$ 100 & 5.50$\pm$ 0.30 & +0.30$\pm$ 0.30 & ... & ... & ... \\*
                             & L2017  & 3720$\pm$ 100 & 4.74$\pm$ 0.10 & +0.20$\pm$ 0.05 & 2.07 & 0.35 & 0.15 \\
       \hline
       \noalign{\smallskip}
       \multirow{5}{*}{11 GJ908}& P2018V  & 3657$\pm$ 51 & 4.84$\pm$ 0.07 & -0.12$\pm$ 0.16 & 3.00 & 0.30 & 0.60\\
                             & P2019V  & 3630$\pm$ 51 & 4.73$\pm$ 0.04 & -0.01$\pm$ 0.16 & 2.00 & 0.30 & 0.60\\
                             & P2019N  & 3651$\pm$ 56 & 4.78$\pm$ 0.04 & -0.22$\pm$ 0.16 & 2.00 & 0.30 & 0.60\\
                             & P2019NV & 3626$\pm$ 54 & 4.73$\pm$ 0.06 & -0.02$\pm$ 0.19 & 2.00 & 0.30 & 0.60\\
                             & R2018NV & 3600$\pm$ 100 & 5.50$\pm$ 0.30 & +0.50$\pm$ 0.30 & ... & ... & ... \\
                             & L2017  & 3646$\pm$ 100 & 4.86$\pm$ 0.10 & -0.51$\pm$ 0.05 & 2.25 & 0.35 & 3.70 \\                     
                    
\end{longtable}
}

\section{Atomic and molecular data}

This appendix provides two tables with the atomic and molecular data in the optical and near-infrared wavelength regions, referred to in Sect.~\ref{sect:method}.

\begin{table*}[ht]
\caption{Atomic and molecular data for selected lines with an estimated central depth larger than 0.01 in a wavelength region around the \ion{K}{I} resonance line at 7699~\AA. All lines are from neutral species.}
\label{tab:lines_7699}
\centering
\begin{tabular}{lrrrrrl}
\hline\hline\noalign{\smallskip}
Species & $\lambda$ & $E_{\rm low}$ & $\log gf$ & $\log\gamma_{\rm rad}$ & $\log\gamma_{\rm Waals}$ & References \\
 & [\AA] & [eV] & & & & \\
\noalign{\smallskip}\hline\noalign{\smallskip}
TiO &       7697.0118 &    0.3741 &   0.041 &  7.020 &  0.000 & \\
TiO &       7697.0118 &    0.4150 &  -0.238 &  6.997 &  0.000 & \\
TiO &       7697.0178 &    0.3141 &  -0.580 &  7.029 &  0.000 & \\
TiO &       7697.0415 &    0.1539 &  -0.723 &  7.016 &  0.000 & \\
TiO &       7697.2015 &    0.2153 &  -0.108 &  7.007 &  0.000 & \\
TiO &       7697.2074 &    0.6242 &  -0.043 &  6.964 &  0.000 & \\
TiO &       7697.5630 &    0.7979 &   0.036 &  6.949 &  0.000 & \\
TiO &       7697.5630 &    0.2903 &  -0.360 &  7.006 &  0.000 & \\
Sc &        7697.7703 &    2.5689 &  -0.410 &  8.190 & -7.710 &    K09 \\
TiO &       7698.1972 &    0.5280 &  -0.018 &  7.003 &  0.000 & \\
TiO &       7698.2091 &    0.4021 &  -0.147 &  6.985 &  0.000 & \\
TiO &       7698.2447 &    0.2202 &  -0.097 &  7.007 &  0.000 & \\
TiO &       7698.2921 &    0.3794 &   0.052 &  7.020 &  0.000 & \\
TiO &       7698.2921 &    0.1568 &  -0.699 &  7.016 &  0.000 & \\
TiO &       7698.4936 &    0.3176 &  -0.558 &  7.028 &  0.000 & \\
TiO &       7698.6656 &    0.4169 &   0.115 &  6.990 &  0.000 & \\
K &         7698.9643 &    0.0000 &  -0.154 &  7.600 & -7.445 & K12, BPM \\
TiO &       7699.3000 &    0.2253 &  -0.086 &  7.006 &  0.000 & \\
TiO &       7699.3178 &    0.6354 &  -0.038 &  6.963 &  0.000 & \\
OH &        7699.4606 &    0.0952 &  -8.073 &  0.000 &  0.000 &    GSGCD \\
TiO &       7699.4838 &    0.4107 &  -0.141 &  6.984 &  0.000 & \\
TiO &       7699.4838 &    0.1598 &  -0.678 &  7.015 &  0.000 & \\
Yb &        7699.4870 &    2.4438 &  -0.034 &  0.000 &  0.000 &    PK \\
TiO &       7699.6320 &    0.5754 &   0.215 &  6.977 &  0.000 & \\
TiO &       7699.6498 &    0.5362 &  -0.012 &  7.002 &  0.000 & \\
TiO &       7699.6498 &    0.3848 &   0.062 &  7.019 &  0.000 & \\
TiO &       7699.6498 &    0.2967 &  -0.351 &  7.005 &  0.000 & \\
TiO &       7699.6617 &    0.4234 &  -0.230 &  6.996 &  0.000 & \\
OH &        7699.9474 &    0.0954 &  -8.073 &  0.000 &  0.000 &    GSGCD \\
TiO &       7699.9760 &    0.3211 &  -0.537 &  7.027 &  0.000 & \\
TiO &       7700.3081 &    0.8108 &   0.040 &  6.947 &  0.000 & \\
Ti &        7700.3312 &    3.1608 &  -1.757 &  7.260 & -7.770 &    K10 \\
TiO &       7700.3852 &    0.2305 &  -0.075 &  7.006 &  0.000 & \\
Ti &        7700.6473 &    3.1608 &  -1.914 &  7.260 & -7.770 &    K10 \\
TiO &       7700.6996 &    0.4255 &   0.122 &  6.989 &  0.000 & \\
TiO &       7700.6996 &    0.1629 &  -0.657 &  7.015 &  0.000 & \\
TiO &       7700.7886 &    0.4193 &  -0.135 &  6.983 &  0.000 & \\
MgH &       7700.8709 &    1.2288 &  -1.377 &  7.060 &  0.000 &    KMGH \\
MgH &       7700.8709 &    1.2288 &  -1.632 &  7.060 &  0.000 &    KMGH \\

\noalign{\smallskip}\hline
\end{tabular}
\tablefoot{$\lambda$ \ldots wavelength, $E_{\rm low}$ \ldots lower level energy, $\log gf$ \ldots logarithm (base 10) of the product of the oscillator strength of the transition and the statistical weight of the lower level, $\log\gamma_{\rm rad}$ \ldots logarithm of the radiative damping width in units of rad~s$^{-1}$, $\log\gamma_{\rm Waals}$ \ldots logarithm of the van der Waals broadening width per unit perturber number density at 10\,000~K in units of rad s$^{-1}$ cm$^{3}$. Unknown damping parameters are set to zero.
Reference for TiO lines: \citet{BP09ow}.
References for other lines:
BPM    \ldots \citet{BPM},
GSGCD  \ldots \citet{GSGCD},
K09    \ldots \citet{K09},
K10    \ldots \citet{K10},
KMGH   \ldots \citet{KMGH},
PK     \ldots \citet{PK},
K12    \ldots \citet{K12}.
}
\end{table*}

\onecolumn
\longtab[2]{
\begin{longtable}{lrrrrrl}
\caption{\label{tab:lines_NIR} Atomic data for selected lines with an estimated central depth larger than 0.15 in wavelength regions around the near-infrared lines investigated in Sects.~\ref{sect:NLTE} and \ref{sec:discussion}. All lines are from neutral species.} \\
\hline\hline\noalign{\smallskip}
Species & $\lambda$ & $E_{\rm low}$ & $\log gf$ & $\log\gamma_{\rm rad}$ & $\log\gamma_{\rm Waals}$ & References \\
 & [\AA] & [eV] & & & & \\
\noalign{\smallskip}\hline\noalign{\smallskip}
\endfirsthead
\caption{Continued.} \\
\hline\hline\noalign{\smallskip}
Species & $\lambda$ & $E_{\rm low}$ & $\log gf$ & $\log\gamma_{\rm rad}$ & $\log\gamma_{\rm Waals}$ & References \\
 & [\AA] & [eV] & & & & \\
\noalign{\smallskip}\hline\noalign{\smallskip}
\endhead
\noalign{\smallskip}\hline\noalign{\smallskip}
\endfoot
\noalign{\smallskip}\hline\noalign{\smallskip}
\multicolumn{7}{l}{\parbox{0.80\textwidth}{For column descriptions see Table~\ref{tab:lines_7699}.
References:
B-WPNP \ldots \citet{B-WPNP},
BPM    \ldots \citet{BPM},
BWL    \ldots \citet{BWL},
GSGCD  \ldots \citet{GSGCD},
K07    \ldots \citet{K07},
K08    \ldots \citet{K08},
K09    \ldots \citet{K09},
K10    \ldots \citet{K10},
K12    \ldots \citet{K12},
K12    \ldots \citet{K12},
K14    \ldots \citet{K14},
KP     \ldots \citet{KP},
LGWSC  \ldots \citet{LGWSC},
MFW    \ldots \citet{MFW},
NIST10 \ldots \citet{NIST10},
NIST10 \ldots \citet{NIST10},
WSM    \ldots \citet{WSM},
WSM    \ldots \citet{WSM},
WV     \ldots \citet{WV}.
}}\\
\endlastfoot
Cr &         11015.530 &   3.4493 &  -0.429 &  8.370 & -7.530 &  K10 \\
K &          11019.848 &   2.6700 &  -0.010 &  7.540 & -6.661 &  WSM, BPM \\
K &          11022.653 &   2.6702 &  -0.161 &  7.560 & -6.661 &  K12, BPM \\
Cr &         11044.610 &   3.0111 &  -1.930 &  6.980 & -7.780 &  K10 \\
OH &         11066.781 &   0.7819 &  -5.947 &  0.000 &  0.000 &  GSGCD \\
OH &         11066.781 &   0.7819 &  -5.947 &  0.000 &  0.000 &  GSGCD \\
OH &         11068.946 &   0.7826 &  -5.948 &  0.000 &  0.000 &  GSGCD \\
OH &         11068.946 &   0.7826 &  -5.948 &  0.000 &  0.000 &  GSGCD \\
Fe &         11607.572 &   2.1979 &  -2.009 &  7.160 & -7.820 &  BWL \\
Cr &         11610.560 &   3.3212 &   0.055 &  7.850 & -7.640 &  K10 \\
Fe &         11638.260 &   2.1759 &  -2.214 &  7.170 & -7.820 &  BWL \\
Fe &         11689.972 &   2.2227 &  -2.068 &  7.150 & -7.820 &  BWL \\
K &          11690.220 &   1.6100 &   0.250 &  7.810 & -7.326 &  WSM, BPM \\
Ca &         11759.570 &   4.5313 &  -0.878 &  7.500 & -7.090 &  K07 \\
Ca &         11767.481 &   4.5322 &  -0.536 &  7.500 & -7.090 &  K07 \\
Ca &         11769.345 &   4.5322 &  -1.011 &  7.500 & -7.090 &  K07 \\
K &          11769.639 &   1.6171 &  -0.450 &  7.810 & -7.326 &  WSM, BPM \\
K &          11772.838 &   1.6171 &   0.510 &  7.810 & -7.326 &  WSM, BPM \\
Ti &         11780.542 &   1.4432 &  -2.170 &  6.870 & -7.790 &  LGWSC \\
Fe &         11783.265 &   2.8316 &  -1.574 &  6.750 & -7.820 &  BWL \\
Ca &         11793.043 &   4.5347 &  -0.258 &  7.500 & -7.090 &  K07 \\
Ca &         11795.763 &   4.5347 &  -1.008 &  7.500 & -7.090 &  K07 \\
Ti &         11797.186 &   1.4298 &  -2.280 &  6.900 & -7.790 &  LGWSC \\
Mg &         11828.171 &   4.3458 &  -0.333 &  0.000 & -7.192 &  NIST10, BPM \\
Fe &         11882.844 &   2.1979 &  -1.668 &  7.170 & -7.820 &  BWL \\
Fe &         11884.083 &   2.2227 &  -2.083 &  7.160 & -7.820 &  BWL \\
Ti &         11892.877 &   1.4298 &  -1.730 &  6.930 & -7.790 &  LGWSC \\
V &          11911.885 &   2.3654 &  -0.874 &  6.340 & -7.780 &  K09 \\
Ti &         11949.547 &   1.4432 &  -1.570 &  6.900 & -7.790 &  LGWSC \\
Ca &         11955.955 &   4.1308 &  -0.849 &  8.010 & -7.300 &  K07 \\
Fe &         11973.046 &   2.1759 &  -1.483 &  7.190 & -7.820 &  BWL \\
Ti &         11973.847 &   1.4601 &  -1.390 &  6.870 & -7.790 &  LGWSC \\
Mg &         12083.278 &   5.7532 &   0.450 &  0.000 &  0.000 &  KP \\
Mg &         12083.649 &   5.7532 &   0.410 &  7.470 & -6.981 &  NIST10 \\
Ca &         12105.841 &   4.5541 &  -0.305 &  7.420 & -7.090 &  K07 \\
Fe &         12190.098 &   3.6352 &  -2.330 &  8.070 & -7.750 &  BWL \\
Na &         12311.480 &   3.7526 &  -1.007 &  0.000 &  0.000 &  NIST10 \\
Na &         12319.980 &   3.7533 &  -0.753 &  0.000 &  0.000 &  NIST10 \\
K &          12432.273 &   1.6100 &  -0.439 &  7.790 & -7.022 &  WSM, BPM \\
Ca &         12433.748 &   5.0261 &  -0.066 &  7.940 & -7.090 &  K07 \\
Ti &         12484.617 &   1.5025 &  -3.277 &  6.870 & -7.780 &  K10 \\
Cr &         12521.810 &   2.7079 &  -1.587 &  7.290 & -7.800 &  K10 \\
K &          12522.134 &   1.6171 &  -0.139 &  7.790 & -7.021 &  WSM, BPM \\
Cr &         12532.840 &   2.7088 &  -1.879 &  7.290 & -7.800 &  K10 \\
Fe &         12556.996 &   2.2786 &  -3.626 &  7.160 & -7.820 &  BWL \\
Ti &         12569.571 &   2.1747 &  -2.050 &  6.540 & -7.810 &  LGWSC \\
Ti &         12600.277 &   1.4432 &  -2.320 &  6.810 & -7.790 &  LGWSC \\
Ca &         12610.942 &   5.0486 &  -0.063 &  8.020 & -6.770 &  K07 \\
Fe &         12638.703 &   4.5585 &  -0.783 &  8.440 & -7.540 &  K14 \\
Fe &         12648.741 &   4.6070 &  -1.140 &  8.420 & -7.540 &  BWL \\
Ti &         12671.096 &   1.4298 &  -2.360 &  6.820 & -7.790 &  LGWSC \\
Na &         12679.170 &   3.6170 &  -0.043 &  0.000 & -6.653 &  NIST10, BPM \\
Na &         12679.170 &   3.6170 &  -1.344 &  0.000 & -6.653 &  NIST10, BPM \\
Na &         12679.220 &   3.6170 &  -0.197 &  0.000 & -6.653 &  NIST10, BPM \\
Ti &         12738.383 &   2.1747 &  -1.280 &  7.950 & -7.750 &  LGWSC \\
Ti &         12744.905 &   2.4875 &  -1.280 &  7.530 & -7.770 &  LGWSC \\
Fe &         12807.152 &   3.6398 &  -2.452 &  8.080 & -7.750 &  K14 \\
Ti &         12811.478 &   2.1603 &  -1.390 &  7.990 & -7.750 &  LGWSC \\
Ca &         12816.045 &   3.9104 &  -0.765 &  8.280 & -7.520 &  K07 \\
Ti &         12821.672 &   1.4601 &  -1.190 &  6.810 & -7.790 &  LGWSC \\
Ca &         12823.867 &   3.9104 &  -0.997 &  8.280 & -7.520 &  K07 \\
Ca &         12827.059 &   3.9104 &  -1.478 &  8.280 & -7.520 &  K07 \\
Ti &         12831.445 &   1.4298 &  -1.490 &  6.820 & -7.790 &  LGWSC \\
Ti &         12847.034 &   1.4432 &  -1.330 &  6.820 & -7.790 &  LGWSC \\
Fe &         12879.766 &   2.2786 &  -3.458 &  7.170 & -7.820 &  BWL \\
Ca &         12885.290 &   4.4300 &  -1.164 &  7.770 & -7.710 &  K07 \\
Mn &         12899.760 &   2.1142 &  -1.070 &  0.000 &  0.000 &  B-WPNP \\
V &          12901.212 &   1.9553 &  -1.052 &  6.890 & -7.780 &  K09 \\
Ca &         12909.070 &   4.4300 &  -0.224 &  7.770 & -7.710 &  K07 \\
Cr &         12910.090 &   2.7079 &  -1.779 &  7.260 & -7.800 &  K10 \\
Ti &         12919.899 &   2.1535 &  -1.560 &  8.000 & -7.750 &  LGWSC \\
Cr &         12921.810 &   2.7088 &  -2.743 &  7.260 & -7.800 &  K10 \\
Ni &         12932.313 &   2.7403 &  -2.523 &  7.680 & -7.810 &  K08 \\
Cr &         12937.020 &   2.7099 &  -1.896 &  7.260 & -7.800 &  K10 \\
Mn &         12975.910 &   2.8884 &  -1.090 &  0.000 &  0.000 &  B-WPNP \\
Ti &         12987.567 &   2.5057 &  -1.550 &  7.530 & -7.770 &  LGWSC \\
Ca &         13001.402 &   4.4410 &  -1.139 &  7.770 & -7.710 &  K07 \\
Ti &         13005.365 &   2.1747 &  -2.287 &  7.990 & -7.750 &  K10 \\
Fe &         13006.684 &   2.9904 &  -3.744 &  6.120 & -7.810 &  K14 \\
Ti &         13011.250 &   2.1603 &  -2.180 &  8.000 & -7.750 &  LGWSC \\
Ti &         13011.897 &   1.4432 &  -2.270 &  6.820 & -7.790 &  LGWSC \\
Ca &         13033.554 &   4.4410 &  -0.064 &  7.770 & -7.710 &  K07 \\
Ca &         13057.885 &   4.4410 &  -1.092 &  7.770 & -7.710 &  K07 \\
Ti &         13077.265 &   1.4601 &  -2.220 &  6.820 & -7.790 &  LGWSC \\
Ca &         13086.430 &   4.4430 &  -1.214 &  7.810 & -7.690 &  K07 \\
V &          13104.517 &   1.9496 &  -1.238 &  6.890 & -7.780 &  K09 \\
Al &         13123.410 &   3.1427 &   0.270 &  0.000 &  0.000 &  WSM \\
Ca &         13134.942 &   4.4506 &   0.085 &  7.770 & -7.710 &  K07 \\
Al &         13150.753 &   3.1427 &  -0.030 &  0.000 &  0.000 &  WSM \\
Ca &         13167.759 &   4.4506 &  -1.092 &  7.770 & -7.710 &  K07 \\
Cr &         13201.150 &   2.7088 &  -1.834 &  7.250 & -7.800 &  K10 \\
Cr &         13217.020 &   2.7099 &  -2.302 &  7.250 & -7.800 &  K10 \\
Ca &         13250.322 &   4.5541 &  -1.033 &  7.560 & -7.100 &  K07 \\
Ti &         13255.812 &   2.2312 &  -2.119 &  6.260 & -7.810 &  K10 \\
Mn &         13281.490 &   2.9197 &  -1.350 &  0.000 &  0.000 &  B-WPNP \\
Fe &         13287.829 &   2.9488 &  -3.021 &  6.130 & -7.810 &  BWL \\
V &          13291.120 &   1.9452 &  -1.570 &  0.000 &  0.000 &  MFW \\
V &          13291.285 &   1.9452 &  -1.406 &  6.880 & -7.780 &  K09 \\
Mn &         13293.800 &   2.1427 &  -1.580 &  0.000 &  0.000 &  B-WPNP \\
Ti &         13305.697 &   2.2393 &  -1.863 &  6.080 & -7.810 &  K10 \\
Ca &         13317.984 &   4.6244 &  -0.480 &  7.660 & -7.130 &  K07 \\
Mn &         13318.940 &   2.1427 &  -1.370 &  0.000 &  0.000 &  B-WPNP \\
Ti &         13346.704 &   2.2497 &  -2.243 &  5.820 & -7.800 &  K10 \\
Lu &         13371.782 &   0.0000 &  -1.000 &  0.000 &  0.000 &  WV \\
Fe &         13392.102 &   5.3516 &  -0.125 &  8.220 & -7.480 &  K14 \\
Mn &         15159.158 &   4.8889 &   0.619 &  8.040 & -7.520 &  K07 \\
K &          15163.067 &   2.6700 &   0.689 &  7.640 & -7.320 &  K12 \\
K &          15163.067 &   2.6700 &  -0.613 &  7.640 & -7.320 &  K12 \\
K &          15168.376 &   2.6702 &   0.480 &  7.620 &  0.000 &  WSM \\

\end{longtable}
}

\end{appendix}
\end{document}